\documentclass[aps,pra,twocolumn,superscriptaddress,nofootinbib]{revtex4-2}

\usepackage{graphicx}
\usepackage{latexsym}
\usepackage{amsmath}
\usepackage{amssymb}
\usepackage{amsfonts}
\usepackage{color}
\usepackage{pgf}
\usepackage{amsthm}
\usepackage{dsfont}
\usepackage{tikz}
\usetikzlibrary{matrix}

\usepackage[utf8]{inputenc}
\usepackage[english]{babel}
\usepackage{ulem}
\usepackage{mwe}
\usepackage{mathtools}
\usepackage{amssymb}
\usepackage{colortbl}
\begin{document}

\newcommand\mean[1]{\langle #1 \rangle}
\newcommand\dmean[1]{\langle \langle #1 \rangle \rangle}
\newcommand\hatd[1]{\hat{#1}^{\dagger}}
\newcommand\hatdn[2]{\hat{#1}^{\dagger #2}}
\newcommand\hatn[2]{\hat{#1}^{#2}}
\newcommand\matthree[9]{\left(\begin{array}{c c c} #1 & #2 & #3 \\ #4 & #5 & #6 \\ #7 & #8 & #9\end{array}\right)}
\newcommand{\ket}[1]{\vert #1 \rangle}
\newcommand{\bra} [1] {\langle #1 \vert}
\newcommand{\braket}[2]{\langle #1 | #2 \rangle}
\newcommand{\ketbra}[2]{| #1 \rangle \langle #2 |}
\newcommand{\proj}[1]{\ket{#1}\bra{#1}}
\newcommand{\norm}[1]{|#1|}
\newcommand{\opnorm}[1]{|\!|\!|#1|\!|\!|_2}
 \newcommand{\kket}[1]{\vert\vert #1 \rangle\rangle}
 \newcommand{\bbra} [1] {\langle \langle \rangle#1 \vert\vert}
\newcommand{\mmean}[1]{\langle\langle #1 \rangle\rangle}
\newcommand{\tr}{\mathrm{Tr}}
\newcommand{\red}[1]{\textcolor{red}{#1	}}
\newcommand{\blue}[1]{\textcolor{blue}{#1	}}

\newtheoremstyle{break}
  {\topsep}{\topsep}%
  {\itshape}{}%
  {\bfseries}{}%
  {\newline}{}%
\theoremstyle{break}
\newtheorem{theorem}{Theorem}
\newtheorem{lem}{Lemma}
\newtheorem{rem}{Remark}
\newtheorem{defin}{Definition}
\newtheorem{corollary}{Corollary}
 \newtheorem{conj}{Conjecture}
 \newtheorem*{prop}{Properties}

\title{Multicopy observables for the detection of optically nonclassical states}

\author{Matthieu Arnhem}
\author{Célia Griffet}
\author{Nicolas J. Cerf}
\affiliation{Centre for Quantum Information and Communication, \'Ecole polytechnique de Bruxelles, CP 165, Universit\'e libre de Bruxelles, 1050 Brussels, Belgium}

\begin{abstract}
Distinguishing quantum states that admit a classical counterpart from those that exhibit nonclassicality has long been a central issue in quantum optics. Finding an implementable criterion certifying optical nonclassicality (i.e, the incompatibility with a statistical mixture of coherent states) is of major importance as it often is a prerequisite to quantum information processes. A hierarchy of conditions for detecting whether a quantum state exhibits optical nonclassicality can be written based on matrices of moments of the optical field [E. V. Shchukin and W. Vogel, Phys. Rev. A 72, 043808 (2005)]. Here, we design optical nonclassicality observables that act on several replicas of a quantum state and whose expectation value coincides with the determinant of these matrices, hence providing witnesses of optical nonclassicality that overcome the need for state tomography. These multicopy observables are used to construct a family of physically implementable schemes involving linear optical operations and photon number detectors. 
\end{abstract}

\maketitle

\nopagebreak

\section{Introduction}\label{sec:introduction}

Determining whether a quantum optical state admits nonclassical properties or not is an ubiquitous question in the theory of quantum optics as well as in the development of quantum technologies. Numerous proposals for identifying quantum states displaying nonclassicality have been discussed in the literature, see e.g. \cite{agarwal_nonclassical_1992,lutkenhaus_nonclassical_1995, richter_nonclassicality_2002, rivas_nonclassicality_2009, ryl_unified_2015, PhysRevLett.122.080402, PhysRevA.102.032413} or see \cite{miranowicz_testing_2010} for a review. We focus here on a possible definition of optical nonclassicality as introduced by Glauber and Sudarshan \cite{glauber_coherent_1963, glauber_quantum_1963, sudarshan_equivalence_1963}, starting from the assertion that classical states are those that are expressible as convex mixtures of coherent states. Accordingly, when the Glauber-Sudarshan $P$ function of a state is incompatible with a true probability distribution (i.e., when it admits negative values in phase space or is not well-behaved in the sense that it cannot be expressed as a regular function), the state is said to be optically nonclassical. A straightforward operational meaning of optical nonclassicality is that it is a necessary condition in order to produce entanglement with a beam splitter \cite{kim_entanglement_2002}. More generally, being able to identify and characterize such nonclassical optical states is essential since nonclassical features are often taken as resources for quantum information tasks \cite{obrien_photonic_2009, slussarenko_photonic_2019} such as quantum computation \cite{knill_scheme_2001}, distributed quantum computing \cite{shahandeh_quantum_2019}, quantum networks \cite{yadin_operational_2018}, quantum boson sampling \cite{shahandeh_quantum_2017}, quantum metrology \cite{kwon_nonclassicality_2019,Ge_ressource_theory_2020} or quantum communication \cite{ralph_bright_2009}.

Various implementation methods have been proposed for identifying nonclassical states, exploiting measurements ranging from single-photon detection \cite{sperling_uncovering_2015, avenhaus_accessing_2010} to continuous-variable measurements such as homodyne detection \cite{shchukin_nonclassical_2005} or heterodyne detection \cite{bohmann_probing_2020}. In this paper, we introduce a technique that uses multiple replicas (i.e., identical copies) of a quantum state in order to construct a nonclassicality observable. The relevance of multicopy observables in quantum optics has recently been explored in the context of uncertainty relations \cite{hertz_multicopy_2019}. It relies on the observation that polynomial functions of the elements of a density matrix can be expressed by defining an observable acting on several replicas of the state, avoiding the need for quantum tomography \cite{brun_measuring_2004}.  In the present paper, we consider the nonclassicality criteria resulting from the matrix of moments of the optical field as introduced in \cite{shchukin_nonclassicality_2005,shchukin_nonclassical_2005}. In the simplest cases at hand, the multicopy observables enable an original implementation of nonclassicality witnesses through linear interferometry and photon number detectors.

This paper is constructed as follows. In Sec. \ref{sec:nonclassicality}, we review the concept of optical nonclassicality and the link with the matrix of moments of the optical field. In Sec. \ref{sec:performancesmatrixofmoments}, we benchmark the performances of the nonclassicality criteria derived from the determinant of these matrices \cite{shchukin_nonclassical_2005, shchukin_nonclassicality_2005}. We consider a variety of states that are known to be nonclassical (Fock states, squeezed states, cat states, squeezed thermal states) in order to identify which criteria do detect them. This is useful to guide our search for designing multicopy nonclassicality observables, as carried out in Sec. \ref{sec:Implementationmulticopy}. There, we focus on  the most interesting criteria based on a few selected principal minors of the matrix of moments as identified in Sec.~\ref{sec:performancesmatrixofmoments} and provide a physical implementation whenever possible. Finally, in Sec. \ref{sec:Conclusions}, we give our conclusions and discuss further perspectives.

\section{Optical nonclassicality of quantum states}\label{sec:nonclassicality}

Any density operator $\hat{\rho}$ representing the quantum state of a single oscillator (bosonic) mode can be represented in a diagonal form in the coherent state basis  $\vert \alpha \rangle$, namely
\begin{equation}\label{Def:Pfunction}
\hat{\rho} = \int P(\alpha) \vert \alpha \rangle \langle \alpha \vert d^2 \alpha,
\end{equation}
where $P(\alpha)$ is the Glauber-Sudarshan $P$ function. Note that $P(\alpha)$ completely defines state $\hat{\rho}$ and is normalized since Tr$(\hat{\rho})=1$.  
A state $\hat{\rho}$ is said to be classical if its associated $P$ function behaves as a probability distribution $P(\alpha) = P_{cl}(\alpha)$, hence it is non-negative. Any convex mixture of coherent states $\ket{\alpha}$ is thus classical by definition.
Conversely, a quantum state $\hat{\rho}$ is considered to be nonclassical if it cannot be written as a mixture of coherent states, i.e., if $P(\alpha) \neq P_{cl}(\alpha)$. Simple examples of nonclassical states include Fock states or squeezed states, whose $P$ functions are not regular (their expressions involve derivatives of Dirac $\delta$-functions).

The expectation value of any normally-ordered operator function $:\! {\hat g}(\hat{a}, \hatd{a}) \!:$ of the annihilation $\hat{a}$ and creation $\hatd{a}$ operators can be expressed using the $P$ function as
\begin{equation}\label{def:expectationvaluePfunction}
\langle \,:\! {\hat g}(\hat{a}, \hatd{a}) \! :\, \rangle = \int d^2 \alpha \, P(\alpha) \, g(\alpha, \alpha^*).
\end{equation}
In this expression, the vertical pair of dots stands for normal ordering, which means that all creation operators must be placed on the left of annihilation operators. 
Hence, if the $P$ function $P(\alpha)$ admits negative values, then Eq.~(\ref{def:expectationvaluePfunction}) can become negative for some well chosen function ${\hat g}(\hat{a}, \hatd{a})$, witnessing the nonclassicality of state $\hat{\rho}$. This suggests a close connection between the expectation value of normally-ordered functions and the nonclassical character of the $P$ function: as observed in Ref. \cite{shchukin_nonclassicality_2005}, any normally-ordered Hermitian operator of the form $:\!\!\hat{f}^{\dagger} \hat{f}\!\!:$ can be used to detect nonclassicality. The expectation value of $:\!\!\hat{f}^{\dagger} \hat{f}\!\!:$ can indeed be written in terms of the $P$ function as 
\begin{equation}
\mean{\,:\!\!\hat{f}^{\dagger} \hat{f} \!\!:\,} = \int d^2 \alpha \, P(\alpha) \, \vert f(\alpha) \vert^2,
\end{equation}
which is always positive for any $f(\alpha)$ provided $P(\alpha)$ is a classical probability distribution $P_{cl}(\alpha)$. Hence, a witness of nonclassicality of the considered state $\hat{\rho}$ is provided by the existence of negative expectation values for some suitably chosen operator function $\hat{f}$, that is,
\begin{equation}\label{eq:nonclassicalitycriteria}
\exists \,\hat{f} \qquad \mathrm{s.t.}\qquad \mean{ \,:\!\! \hat{f}^{\dagger} \hat{f}\!\!:\, } < 0.
\end{equation}

Furthermore, as shown in Ref. \cite{shchukin_nonclassicality_2005}, these nonclassicality  criteria can be reformulated in terms of an infinite countable set of inequalities, which involve the principal minors of an infinite-dimensional matrix of moments. The infinite set of inequalities completely characterizes the nonclassicality of the quantum state under study. As shown in Ref. \cite{shchukin_nonclassical_2005}, these criteria can be constructed for three different sets of operators $(\hat{a},\hat{a}^\dagger)$, $(\hat{x}_\phi,\hat{p}_\phi)$ and $(\hat{x}_ \phi,\hat{n})$, but we only consider the set $(\hat{a},\hat{a}^\dagger)$ in the present paper. In this case, one exploits the fact that any operator $\hat{f}$ can be expressed as a (normally-ordered) Taylor series \begin{equation}
	 \hat{f}(\hat{a},\hatd{a}) = \sum_{k=0}^{\infty} \sum_{l=0}^{\infty} c_{kl} ~ \hat{a}^{\dagger k} \, \hat{a}^l .
\end{equation}
Hence, a necessary criterion for classicality can be reformulated as
\begin{equation}
	\langle \, : \!\! \hat{f}^\dagger \hat{f} \!\! : \, \rangle =\sum_{k,m=0}^{\infty} \sum_{l,n=0}^{\infty} c^*_{mn} c_{kl}  \, \langle \hat{a}^{\dagger (k+n)} \, \hat{a}^{l+m}\rangle \geq 0 ,
\end{equation}
for any coefficients $c_{ij}$'s. By using Silvester's criterion, this can be reexpressed as the positivity of the determinant of the matrix of moments $D_N$, defined as
\begin{widetext}

\begin{equation}\label{def:dNmatrix}
    D_N =
    \begin{pmatrix}
        1 & \langle \hat{a} \rangle & \langle \hatd{a} \rangle & \langle \hat{a}^2 \rangle & \langle \hatd{a} \hat{a} \rangle & \langle   \hatdn{a}{2}   \rangle & ...\\
        \langle \hatd{a} \rangle & \langle \hatd{a} \hat{a} \rangle & \langle \hat{a}^{\dagger 2} \rangle & \langle \hatd{a} \hat{a}^2 \rangle & \langle \hat{a}^{\dagger 2} \hat{a} \rangle & \langle   \hatdn{a}{3}   \rangle & ...\\
        \langle \hat{a} \rangle & \langle \hat{a}^2 \rangle & \langle \hatd{a} \hat{a} \rangle & \langle \hat{a}^3 \rangle & \langle \hatd{a} \hat{a}^2 \rangle & \langle   \hatdn{a}{2} \hat{a}   \rangle & ... \\
        \langle \hat{a}^{\dagger 2} \rangle & \langle \hat{a}^{\dagger 2} \hat{a} \rangle & \langle \hat{a}^{\dagger 3} \rangle & \langle \hat{a}^{\dagger 2} \hat{a}^2 \rangle & \langle \hat{a}^{\dagger 3} \hat{a} \rangle & \langle   \hatdn{a}{4}   \rangle & ... \\
        \langle \hatd{a} \hat{a} \rangle & \langle \hatd{a} \hat{a}^2 \rangle & \langle \hat{a}^{\dagger 2} \hat{a} \rangle & \langle \hatd{a} \hat{a}^3 \rangle & \langle \hat{a}^{\dagger 2} \hat{a}^2 \rangle & \langle   \hatdn{a}{3} \hat{a}   \rangle &...\\
        \langle   \hat{a}^2   \rangle & \langle   \hat{a}^3   \rangle & \langle    \hatd{a} \hat{a}^2   \rangle & \langle   \hat{a}^4   \rangle  & \langle   \hatd{a} \hat{a}^3   \rangle & \langle   \hatdn{a}{2} \hat{a}^2   \rangle & ... \\
        \vdots &  \vdots & \vdots & \vdots  & \vdots & \vdots  & \ddots \\
    \end{pmatrix},
\end{equation}
\end{widetext}
which contains all normally-ordered moments of $\hat{a}$ and $\hatd{a}$ up to order $N$. The matrix of moments $D_N$ can be defined for any dimension $N \times N$ (its block structure is discussed in Appendix \ref{app:matrixofmomentsproperties}) and its determinant will be written $d_{1\cdots N} = \det(D_N)$ in the rest of this paper, where the index of $d$ means that all rows and columns of $D_N$ are kept in the range from 1 to $N$. Remarkably, as a consequence of Bochner's theorem, the classicality criteria become necessary and sufficient when the determinants are positive for all orders \cite{shchukin_nonclassical_2005}, that is, $\hat{\rho}$ is classical if and only if
\begin{equation}\label{eq:CNSclassicality}
	d_{1...N} \geq 0,  \space ~ \forall N.
\end{equation}
The determinants $d_{1...N}$ are the so-called \textit{dominant} principal minors of matrix $D_N$, i.e., the determinants of the matrices constructed by taking \textit{all} rows and columns in the upper-left corner of the matrix. Hence, the negativity of any single determinant $d_{1...N}$ of order $N$ is a sufficient condition for nonclassicality.

Note that one can construct various matrices of moments having similar properties and nonclassicality detection power. In what follows, we will focus on the principal minors of the matrix of moments $D_N$, which are built by selecting some rows and corresponding columns and then taking the determinant of the resulting matrix. For example, if rows and columns $i$, $j$, and $k$ are selected, the associated principal minor is written $d_{ijk}$. Interestingly, any principal minor such as $d_{ijk}$ provides a sufficient criterion for nonclassicality: if $d_{ijk}<0$, then the state $\hat{\rho}$ is nonclassical. Some examples of principal minors that are not dominant are $d_{14}, d_{15}, d_{124}, d_{134}$, and $d_{145}$, while examples of dominant principal minors are $d_{12}, d_{123}, d_{1234}$, and $d_{1235}$ (note that we adopt a slightly relaxed definition of dominant principal minors, see \footnote{From the block structure of the matrix of moments $D_N$ as described in Appendix \ref{app:matrixofmomentsproperties}, we slightly adapt the definition of a dominant principal minor: it is associated with the upper left submatrix but assuming the order of rows and columns is irrelevant within a given block. For example, $d_{1235}$ is understood as a dominant principal minor although the fourth row and column are omitted.\label{fn_dominant_minor} }). Each of these minors might have a distinct physical interpretation and hence, detect different types of nonclassical states (see Ref. \cite{miranowicz_testing_2010} for a review of nonclassicality criteria).

In Sec. \ref{sec:performancesmatrixofmoments}, we will study all the principal minors of the matrix of moment $D_N$ of dimension $5 \times 5$, which is helpful to determine which of them are good candidates for constructing a multicopy observable (as will then be addressed in Sec. \ref{sec:Implementationmulticopy}). Before coming to this, it is relevant to list the important properties of the matrix of moments $D_N$ and its determinants (more details and proofs are given in Appendix \ref{app:matrixofmomentsproperties}): 

\begin{itemize}
\item The matrix of moment $D_N$ is Hermitian. Hence, its determinant as well as all its principal minors are real-valued. 

\item All principal minors of matrix $D_N$ vanish for coherent states $\vert \alpha \rangle$. This property is consistent with the fact that the coherent states are on the boundary of the convex set of classical states. Any statistical mixture of coherent states is classical and can only give a higher value of all principal minors. Conversely, a slight deviation from a coherent state to a nonclassical state may induce some principal minor to have a negative value.

\item All principal minors of matrix $D_N$ are invariant under rotations in phase space. Hence, all corresponding nonclassicality criteria are invariant under rotations in phase space. This property is consistent with the fact that  nonclassicality is a feature that is unaffected by such rotations. This simplifies the calculations since all phase terms can be given arbitrary values and will typically be set to zero. 

\item All \textit{dominant} principal minors of matrix $D_N$ are invariant under displacements in phase space. This property is consistent with the fact that nonclassicality is unaffected by such displacements. Hence, we can simplify our calculations by considering states that are centered in phase space. Note that, unfortunately, the \textit{nondominant} principal minors do not enjoy this invariance property. In Section \ref{ssubsec:twomodeinterpolation+displacement}, we will consider the effect of displacements on some non-dominant principal minors and illustrate how it affects the detection capability of the corresponding criteria.    
\end{itemize}

\section{Nonclassicality criteria based on the matrix of moments}
\label{sec:performancesmatrixofmoments}

Let us benchmark the performance of the nonclassicality criteria derived from the principal submatrices of the matrix of moments $D_N$ (up to $N=5$) in terms of their ability to detect various nonclassical states. We express the corresponding principal minors for common classes of nonclassical pure states, such as Fock states, squeezed states, or cat states (note that all these states are centered in phase space). All values are listed in Table \ref{table:determinantdeterminants}, where negative values imply the actual detection of nonclassicality (see entries with gray background). We then study the performance of the criteria when applied to Gaussian (pure or mixed) states and determine that $d_{123}$ is a necessary and sufficient nonclassicality criterion for these states.  Overall, our observations lead us to focus on $d_{15}$, $d_{23}$, $d_{123}$ and $d_{1235}$ when constructing multicopy nonclassicality observables in Sec. \ref{sec:Implementationmulticopy}.

\subsection{Nonclassical pure states}\label{ssubsec:statesclandnoncl}
\vspace{-2mm}
\subsubsection{Fock states} \label{sec:fockstates}
\vspace{-2mm}
Fock states $\ket{n}$ (except the vacuum state $\ket{0}$) are common nonclassical states, which can be detected by criteria such as $d_{15}$, $d_{125}$, $d_{135}$, $d_{145}$, or $d_{1235}$, as can be seen in Table \ref{table:determinantdeterminants}. In a nutshell, the Fock states are only detected when an odd number of off-diagonal entries of the type $\mean{\hatdn{a}{k}\hatn{a}{k}}$ appear in the principal submatrix of the matrix of moments $D_N$. This explains why Fock states are only detected in Table \ref{table:determinantdeterminants} for criteria that involve the fifth row or column since the first non-zero off-diagonal element of $D_5$ is $(D_5)_{1,5} = (D_5)_{5,1}$. For completeness, we list all nonzero moments (entries of matrix $D_5$) in Table \ref{table:tableofmoments}.
\vspace{-2mm}

\subsubsection{Squeezed states}
\vspace{-2mm}
Squeezed states are nonclassical quantum states which can be used, for instance, to enhance the sensitivity of the LIGO experiment \cite{aasi_enhanced_2013,chua_quantum_2014}. In order to check which principal minors detect them as nonclassical states, we need to evaluate the entries of the matrix of moments $D_5$. First, we observe that all moments $\mean{\hatdn{a}{k} \hatn{a}{l}}$ of odd order $k+l$ vanish since squeezed states can be decomposed into even Fock states \cite{weedbrook_gaussian_2012}, that is
\begin{equation}
\vert S_r \rangle = \frac{1}{\sqrt{\cosh r}} \sum_{k=0}^{\infty} (-e^{i \phi} \tanh r)^k \frac{\sqrt{2k}!}{2^k k!} \vert 2 k \rangle,
\end{equation}

\newpage
\onecolumngrid

	\begin{table}[t!]
     $ \begin{aligned}
			\begin{array}{|c|c|c|c|c|c|}
				\hline
				\text{Dimension} & \text{Principal minor} &  \text{Fock states} & \text{Squeezed states} & \text{Odd cat states} & \text{Even cat states}\\ \hline
				2 & d_{12} = d_{13} & n & \sinh^2(r) & |\beta|^2 \frac{N_+}{N_-} & |\beta|^2 \frac{N_-}{N_+} \\  
				2 & d_{14} = d_{16} & n (n-1) & 2 \sinh^4(r) & 0 & 0 \\
				2 & d_{15} & \cellcolor{gray!40} {-n} & \cosh(2r) \sinh^2(r) &\cellcolor{gray!40} |\beta|^4 (1-\frac{N_+^2}{N_-^2}) & |\beta|^4 (1-\frac{N_-^2}{N_+^2}) \\ 
				2 & d_{23} & n^2 & \cellcolor{gray!40}  - \sinh^2(r) & |\beta|^4 (\frac{N_+^2}{N_-^2}-1) & \cellcolor{gray!40} {|\beta|^4 (\frac{N_-^2}{N_+^2}-1)} \\
				2 & d_{24} = d_{26} = d_{34} = d_{36} & n^2 (n-1) & \sinh^4(r) (\cosh^2(r) + 2 \sinh^2(r)) & |\beta|^6  \frac{N_+}{N_-} &  |\beta|^6  \frac{N_-}{N_+}  \\
				2 & d_{25} = d_{35} &  n^2 (n-1) & \sinh^4(r) (\cosh^2(r) + 2 \sinh^2(r)) & |\beta|^6  \frac{N_+}{N_-} &  |\beta|^6  \frac{N_-}{N_+}   \\
				2 & d_{45} = d_{56} & n^2 (n-1)^2 & \cellcolor{gray}{\frac{1}{2} (5 - 3 \cosh(2 r)) \sinh^4(r)} & \cellcolor{gray!40} {|\beta|^8 (1-\frac{N_+^2}{N_-^2})} &  |\beta|^8 (1-\frac{N_-^2}{N_+^2}) \\
				2 & d_{46} & n^2 (n-1)^2 &  \cellcolor{gray!40}{-2 (1 + 3 \cosh(2 r)) \sinh^4(r)} & 0 & 0 \\ \hline
				3 & d_{123} & n^2 & \cellcolor{gray!40} - \sinh^2(r) & |\beta|^4 (\frac{N_+^2}{N_-^2}-1) & \cellcolor{gray!40} |\beta|^4 (\frac{N_-^2}{N_+^2}-1) \\ 
				3 & d_{124} = d_{126} = d_{134} = d_{136} & n^2 (n-1) &  2 \sinh^6(r) & 0 & 0 \\
				3 & d_{125} = d_{135} & \cellcolor{gray!40} -n^2 & \sinh^4(r) \cosh(2r) & \cellcolor{gray!40} |\beta|^6 \frac{N_+}{N_-} (1-\frac{N_+^2}{N_-^2}) & |\beta|^6 \frac{N_-}{N_+} (1-\frac{N_-^2}{N_+^2}) \\ 
				3 & d_{145} = d_{156}  & \cellcolor{gray!40} - n^2(n-1) & \cellcolor{gray!40} - 2 \sinh^6(r) & 0 & 0 \\ 
				3 & d_{146} & n^2 (n-1)^2 & \cellcolor{gray!40} - 4 \cosh(2r) \sinh^4(r) & 0 & 0 \\ 
				3 & d_{234} = d_{236} & n^3 (n-1) & \cellcolor{gray!40} \frac{1}{2} (1-3\cosh(2r)) \sinh(r)^4 & |\beta|^8 (\frac{N_+^2}{N_-^2}-1) & \cellcolor{gray!40} |\beta|^8 (\frac{N_-^2}{N_+^2}-1) \\ 
				3 & d_{235} & n^3 (n-1) & \cellcolor{gray!40}  -\sinh(r)^4 (\cosh(r)^2 + 2 \sinh(r)^2) & |\beta|^8 (\frac{N_+^2}{N_-^2}-1) & \cellcolor{gray!40} {|\beta|^8 (\frac{N_-^2}{N_+^2}-1)} \\ 
				3 & d_{245} = d_{256} = d_{345} = d_{356} & n^3 (n-1)^2 & \cellcolor{gray}  \frac{1}{2} (5-3 \cosh(2r)) \sinh^6(r) & \cellcolor{gray!40} |\beta|^{10} \frac{N_+}{N_-} (1-\frac{N_+^2}{N_-^2}) &  |\beta|^{10} \frac{N_-}{N_+} (1-\frac{N_-^2}{N_+^2}) \\ 
				3 & d_{246} = d_{346} & n^3 (n-1)^2 & \cellcolor{gray!40} - 2 (1+3 \cosh(2r))\sinh^6(r) & 0 & 0 \\ 
				3 & d_{456}  & n^3(n-1)^3 & \cellcolor{gray!40} - 8 \sinh^6(r) & 0 & 0\\ \hline
				4 & d_{1234}  & n^3(n-1)  &  \cellcolor{gray!40} -2 \sinh^6(r) & 0 & 0\\ 
				4 & d_{1235}  & \cellcolor{gray!40} - n^3  &  \cellcolor{gray!40} - \cosh(2r) \sinh^4(r) & \cellcolor{gray!40} - |\beta|^8 (\frac{N_+^2}{N_-^2}-1)^2 & \cellcolor{gray!40} - |\beta|^8 (\frac{N_-^2}{N_+^2}-1)^2 \\ 
				4 & d_{1456} & \cellcolor{gray!40} - n^3(n-1)^2 & \cellcolor{gray!40} - 4 \sinh^6(r) & 0 & 0\\ \hline
				5 & d_{12345} & \cellcolor{gray!40} - n^4(n-1) & 2 \sinh^8(r) & 0 & 0 \\ \hline
			\end{array}
		\end{aligned}
		$
		\caption{Principal minors of the matrix of moments (\ref{def:dNmatrix}) up to dimension N=5 evaluated for different classes of nonclassical (centered) pure states, namely Fock states $\ket{n}$, squeezed states $\ket{S_r}$ of squeezing parameter $r$, and even or odd cat states $\ket{c^\beta_\pm}$ of complex amplitude $\beta$. A light gray background corresponds to states that are detected as nonclassical since the determinant is always negative, while a dark gray background corresponds to states that are only detected as nonclassical above a certain threshold value of the squeezing parameter $r$. The smallest minor that detects Fock states (and odd cat states) is $d_{15}$, while the smallest minor that detects squeezed states (and even cat states) is $d_{23}$. The minor $d_{123}$ is equal to $d_{23}$ for centered states, but, in addition, is invariant under displacements (not visible in the table).
		The minor $d_{1235}$ yields the strongest criterion in this table since it detects Fock, squeezed, and (even and odd) cat states. It is worth mentioning that the minors $d_{24} = d_{26} = d_{34} = d_{36}$ and $d_{25} = d_{35}$ are positive (implying no detection of nonclassicality) for all states for which the moments of odd total order in $\hat{a}$ and $\hatd{a}$ are zero.}
		\label{table:determinantdeterminants}
	\end{table}

	\begin{table}[h!]
		$ \begin{aligned}
			\begin{array}{|c|c|c|c|c|}
				\hline
				\text{Moment} &  \text{Fock} & \text{Squeezed}  & \text{Cat} & \text{Gaussian} \\ \hline
				\mean{\hatd{a} \hat{a}} & n & \sinh^2(r) & |\beta|^2 \frac{N_\mp}{N_\pm} & (\overline{n} + \frac{1}{2}) \cosh(2 r) - \frac{1}{2} \\  
				\mean{\hatn{a}{2}} = \mean{\hatdn{a}{2}}^* & 0 & - \sinh(r) \cosh(r) & \beta^2  & -(\overline{n} + \frac{1}{2}) \sinh(2 r) \\ 
				\mean{\hatdn{a}{2} \hatn{a}{2}} & n(n-1) & \sinh^2(r) \left[ \cosh^2(r) + 2 \sinh^2(r) \right] & |\beta|^4  & \frac{1}{2} (\overline{n} + \frac{1}{2})^2 [3 \cosh(4r)+1] - 2 (\overline{n} + \frac{1}{2}) \cosh(2 r) + 
 \frac{1}{2} \\
				\mean{\hatd{a} \hatn{a}{3}} = \mean{\hatdn{a}{3} \hat{a}}^* & 0 & - 3 \sinh^3(r) \cosh(r) & \beta^2 |\beta|^2 \frac{N_\mp}{N_\pm} & -\frac{3}{2} (\overline{n} + \frac{1}{2})^2 \sinh(4 r) + \frac{3}{2} (\overline{n} + \frac{1}{2}) \sinh(2 r) \\
				\mean{\hatn{a}{4}} = \mean{\hatdn{a}{4}}^*& 0 &  3 \sinh^2(r) \cosh^2(r) & \beta^4   & \frac{3}{2} (\overline{n} + \frac{1}{2})^2 [\cosh(4 r)-1] \\ \hline  
			\end{array}
		\end{aligned}
		$
		\caption{Non-zero moments (entries of the matrix $D_5$) evaluated for different classes of nonclassical (centered) states, namely Fock states $\ket{n}$, squeezed states $\ket{S_r}$ of squeezing parameter $r$, even or odd cat states $\ket{c^\beta_\pm}$ of complex amplitude $\beta$, and Gaussian mixed states (for these, we may consider with no loss of generality a thermal state of mean photon number $\overline{n}$ that is squeezed with a squeezing parameter $r$). All these moments are used to evaluate the nonclassicality criteria based on the minors of the matrix of moments (\ref{def:dNmatrix}) as shown in Tables \ref{table:determinantdeterminants} and \ref{table:determinantgaussian}.}
		\label{table:tableofmoments}
	\end{table}
\clearpage
\twocolumngrid
\noindent where $r$ is the squeezing factor and $\phi$ is the squeezing angle.

The non-vanishing low-order moments are listed in Table \ref{table:tableofmoments} (as a consequence of the rotation invariance, we may  assume $\phi = 0$ without loss of generality). These expressions allow us to easily calculate the different determinants of principal submatrices from the matrix of moments $D_5$ for squeezed states (see Table \ref{table:determinantdeterminants}). The nonclassicality of squeezed states is detected, for example, by $d_{23}$, $d_{123}$, $d_{234}$, $d_{235}$, or $d_{1235}$.

\vspace{-2mm}

\subsubsection{Cat states} \label{sec:catstates}
\vspace{-2mm}
Another archetype of nonclassical states consists of the even and odd optical cat states, written $| c_+ \rangle$ and $|c_- \rangle$ respectively. They are defined as superpositions of coherent states of opposite phases $| \beta \rangle$ and $| -\beta \rangle$, namely 
\begin{align}
	| c_\pm^{\beta} \rangle = \frac{1}{\sqrt{N_\pm}} (| \beta \rangle \pm | -\beta \rangle ),
\end{align}

\noindent where $\beta$ is a complex amplitude and $N_+$ and $N_-$ are normalization constants defined as $N_{\pm} = \sqrt{2 \left(1 \pm e^{- 2 \vert \beta \vert^2}\right)}$. Remember that even cat states and odd cat states are orthogonal to each other, i.e., $ \braket{c_{\pm}^{\alpha}}{{c_{\mp}^{\beta}}} = 0$, while applying the annihilation operator to an odd cat state results in a state proportional to an even cat state and vice-versa:
\begin{equation}\label{eq:catstateannihilationoperator}
\hat{a} \, \ket{c_{\pm}^{\beta}} = \beta \, \sqrt{\frac{N_{\mp}}{N_{\pm}}} \, \ket{c_{\mp}^{\beta}}.
\end{equation}
Therefore, the only non-zero entries in the matrix of moments $D_N$ are those of form $\mean{\hatdn{a}{k} \hatn{a}{l}}$ where $k+l$ is even, as shown in Table \ref{table:tableofmoments} up to $k + l = 4$.
This allows us to calculate the different determinants of principal submatrices from the matrix of moments $D_5$ for even and odd cat states (see Table \ref{table:determinantdeterminants}).

\vspace{-2mm}
\subsubsection{Observations}\label{sec:purestateobservation}
From Table \ref{table:determinantdeterminants}, we can make the following observations (limited to the matrix of moments up to dimension 5) :

\begin{itemize}
\item Increasing the dimension of the matrix of moments does not necessarily lead to a stronger criterion. For example, $d_{12345}$ does not detect more states than $d_{1234}$ while being of higher order. 
\item Some criteria seem to be complementary in the sense that if a state is detected by one criterion, it will not be detected by the complementary criterion and vice-versa. This is, for instance, the case of $d_{15}$ (detecting Fock states but not squeezed states) and $d_{23}$ (detecting squeezed states but not Fock states). Furthermore, it is often the case that a criterion detecting Fock states such as $d_{15}$ also detects odd cat states (similarly, a criterion detecting squeezed states such as $d_{23}$ often detects even cat states). 
\item The strongest criterion seems to be based on $d_{1235}$ since it is the lowest order determinant that detects the nonclassicality of Fock states, squeezed, and (even and odd) cat states.
\end{itemize} 

These observations motivate the rest of this paper, in which we will mostly focus on the multicopy observables for determinants $d_{15}, d_{23}, d_{123}$ and $d_{1235}$.

\bigskip
\subsection{Nonclassical mixed states}

The nonclassicality of mixed states has been studied, for example, in Refs. \cite{kim_properties_1989,lvovsky_nonclassical_2002}. We consider here the simplest case of Gaussian mixed states, for which the limit of nonclassicality is well known: a state is nonclassical if the smallest quadrature variance is smaller than the vacuum noise  variance \cite{kim_properties_1989}. The relevant Gaussian mixed states here are the squeezed thermal states, since a displacement does not affect nonclassicality. It is also sufficient to consider squeezing of the $x$-quadrature since all considered criteria are invariant under rotations. The covariance matrix of these states is written as
\begin{equation}\label{eq:squeezedthermalstatecovariancematrix}
	\gamma^{G} = (\overline{n}+ \frac{1}{2})
	\begin{pmatrix}
		 e^{-2r} & 0 \\
		 0 & e^{2r} \\ 
	\end{pmatrix},
\end{equation}
where $\overline{n}$ is the mean photon number of the thermal state that is squeezed and $r$ is the squeezing parameter.

In order to calculate the principal minors of interest ($d_{15}$, $d_{23}$, $d_{123}$, and $d_{1235}$), we take advantage of the fact that, for Gaussian states, the moments of order higher than two can be expressed as a function of the first- and second-order moments only. Given that squeezed thermal states are centered states, all elements of the matrix of moments $D_N$ can thus be expressed from the covariance matrix $\gamma^{G}$. 
For example, the fourth-order moment $\langle \hat{a}^{\dagger 2} \hat{a}^2 \rangle $ can be calculated from the Wigner function $W(x,p)$ of the state $\hat{\rho}$ by using the overlap formula
\begin{equation}
\langle \hat{A} \rangle= \tr(\hat{A} \hat{\rho}) = \int dx \, dp \, W(x,p) \, \bar{A}(x,p),
\end{equation}
where $\bar{A}(x,p)$ is the Weyl transform of $\hat{A}$. The latter can be obtained by exploiting the commutation relation $[\hat{a},\hat{a}^{\dagger} ]=1$ in such a way as to write $\hat{A}=\hat{a}^{\dagger 2} \hat{a}^2 $ in terms of symmetrically-ordered operators only, namely
\begin{equation}
\begin{split}
  \hat{A} =  S(\hat{a}^{\dagger 2} \hat{a}^2 ) - 2 \, S(\hat{a}^{\dagger} \hat{a} ) + \frac{1}{2},
 \end{split}
\end{equation}
where $S(\cdot)$ denotes symmetric ordering and 
\begin{equation}
\begin{split}
  S(\hat{a}^{\dagger 2} \hat{a}^2 ) &=  \frac{1}{6} ( \hat{a}^{\dagger 2} \hat{a}^2 + \hat{a}^{\dagger} \hat{a}\hat{a}^{\dagger} \hat{a} + \hat{a}^{\dagger} \hat{a}^2 \hat{a}^\dagger + \hat{a} \hat{a}^\dagger \hat{a} \hat{a}^\dagger \\
  & \qquad + \hat{a} \hat{a}^{\dagger 2} \hat{a} +\hat{a}^2 \hat{a}^{\dagger 2} ) ,
  \\
  S(\hat{a}^{\dagger} \hat{a} ) &= \frac{1}{2} (\hat{a}^\dagger \hat{a} + \hat{a} \hat{a}^\dagger).
 \end{split}
\end{equation}
  Hence, $\hat{A}$  can be reexpressed in terms of the $\hat{x}$ and $\hat{p}$ quadrature operators as
\begin{equation}
\begin{split}
  \hat{A} = \frac{1}{12} (\hat{x}^2 \hat{p}^2 + \hat{p}^2 \hat{x}^2 +\hat{x} \hat{p} \hat{x} \hat{p}+\hat{x} \hat{p}^2 \hat{x}+ \hat{p}\hat{x}^2\hat{p}+\hat{p}\hat{x}\hat{p}\hat{x}) \\
 +  \frac{1}{4} (\hat{x}^4 + \hat{p}^4)-(\hat{x}^2 + \hat{p}^2) + \frac{1}{2} .
\end{split}
\end{equation}
The Weyl transform of this expression yields
\begin{equation}
\bar{A}(x,p) = \frac{1}{2} x^2 p^2 + \frac{1}{4} (x^4 + p^4) -(x^2 + p^2) + \frac{1}{2} ,
\end{equation}
so that the mean value of $\hat{A}$ can be written as
\begin{equation}
\begin{split}
 \langle \hat{A} \rangle =   \frac{1}{2} \mean{x^2 p^2} + \frac{1}{4} \left(\mean{x^4} + \mean{p^4}\right) ~~~~~~~~ \\
 -\left(\mean{x^2} + \mean{p^2}\right)
 + \frac{1}{2} .
 \end{split}
 \end{equation}
For a Gaussian distribution, we have
\begin{equation} 
\begin{split}
\langle x^4 \rangle &= 3 \, \Delta x^4 = 3 \, \left(\overline{n} + \frac{1}{2}\right)^2 e^{-4r},\\
\langle x^2 p^2 \rangle &= \Delta x^2 \Delta p^2 =\left(\overline{n} + \frac{1}{2}\right)^2,\\
\langle p^4 \rangle &= 3 \, \Delta p^4 = 3 \, \left(\overline{n} + \frac{1}{2}\right)^2 e^{4r} ,
\end{split}
\end{equation}
so that we get the expression of $\langle \hat{a}^{\dagger 2} \hat{a}^2 \rangle$ that is displayed in Table \ref{table:tableofmoments} for Gaussian states.
By using the same method to calculate the other nonzero moments (also displayed in Table \ref{table:tableofmoments}), we finally obtain the corresponding values of the principal minors shown in Table \ref{table:determinantgaussian}.

\begin{table}[t]
		$ \begin{aligned}
			\begin{array}{|c|c|}
				\hline
				\text{~} &  \text{Squeezed thermal states} \\ \hline
				d_{15}& \frac{1}{4} [1 - 2 (1 + 2 \overline{n}) \cosh(2 r)+ (1 + 2 \overline{n})^2 \cosh(4 r)]  \\ 
				d_{123} = d_{23} &  \frac{1}{2} + \overline{n} + \overline{n}^2 - \frac{1}{2} (1 + 2 \overline{n}) \cosh(2r)  \\  
				d_{1235}  & d_{15} ~ d_{23}  \\ \hline
			\end{array}
		\end{aligned}
		$
		\caption{Principal minors of the matrix of moments $D_5$ evaluated for centered Gaussian states (i.e., squeezed thermal states).}
		\label{table:determinantgaussian}
\end{table}

We see that $d_{15}$ is always positive for any values of parameters $\overline{n}$ and $r$, so it does not yield a criterion. In contrast, $d_{23}$ is interesting as it can be negative for some values of $\overline{n}$ and $r$. Of course, when the mean number of thermal photons $\overline{n} = 0$, we recover the result for squeezed states and $d_{23}=-\sinh^2 (r)$ is negative for all values of the squeezing parameter $r>0$. However, when $\overline{n}>0$, the determinant becomes positive below some threshold value of $r$. From Ref. \cite{kim_properties_1989}, we know that a necessary and sufficient condition for a squeezed thermal state with covariance matrix $\gamma^{G}$ to be nonclassical is
\begin{equation}\label{eq:limitnonclassicald23}
	(\overline{n}+ \frac{1}{2}) \, e^{-2r} < \frac{1}{2}.
\end{equation}
It is easy to check that this precisely corresponds to the condition $d_{23}<0$, so the criterion based on the sign of $d_{23}$ is necessary and sufficient for squeezed thermal states. Note that $d_{23}$ is not invariant under displacements, so that this criterion looses its power when applied to an arbitrary Gaussian state (i.e., a displaced squeezed thermal state). However, the determinant $d_{123}$ can be used instead since it is invariant under displacements and coincides with $d_{23}$ for centered states. Thus, $d_{123}<0$ provides a necessary and sufficient nonclassicality criterion for Gaussian states.

Note finally that for the special case of centered Gaussian states (squeezed thermal states), $d_{1235}$ is equal to the product of $d_{15}$ and $d_{23}$ since all entries of odd order in the annihilation and creation operators in the matrix of moments $D_5$ vanish. Since $d_{15}$ is always positive for Gaussian states, $d_{1235}$ has thus the same detection power as $d_{23}$. Furthermore, since it is invariant under displacements, $d_{1235}$ has the same detection power as $d_{123}$ for arbitrary Gaussian states. Yet, going beyond Gaussian states, we have seen that $d_{1235}$ actually surpasses $d_{123}$ in the sense that it also detects (displaced) Fock states.

This is also true for non-Gaussian mixed states. For example, we may consider a restricted class of non-Gaussian mixed states that are nonclassical as a result of photon addition or subtraction from a Gaussian state. Such states are viewed as essential for quantum computing, see e.g. Refs. \cite{chabaud_classical_2021,walschaers_non-gaussian_2021}. Interestingly, all the moments appearing in the matrix of moments $D_N$ for these states can be deduced from the moments of the corresponding Gaussian states as displayed in Table \ref{table:tableofmoments}. Hence, we can easily derive the principal minors for photon-added and photon-subtracted Gaussian states. It is worth noticing that even the simpler criterion based on $d_{15}$ can detect photon-subtracted Gaussian states for small values of $\overline{n}$ and $r$, while it is useless in the case of Gaussian states.

\section{Multicopy nonclassicality observables}\label{sec:Implementationmulticopy}

Potential implementations of nonclassicality criteria based on the matrix of moments $D_N$ have been considered in Ref. \cite{shchukin_nonclassical_2005}, but the idea was to experimentally evaluate each individual entry of the matrix before calculating its determinant. Instead, in the present work, we look for an optical implementation that makes it possible to directly access the value of the determinant by measuring the expectation value of some nonclassicality observable. Since the principal minors discussed in Sec. \ref{sec:performancesmatrixofmoments}
(especially $d_{15}, d_{23}, d_{123}$, and $d_{1235}$) are polynomial functions of the matrix elements of $\hat{\rho}$, we turn to multicopy observables as defined in Ref. \cite{brun_measuring_2004}. This method appears to be well adapted here since the nonclassicality criteria involve determinants (a similar technique has been successfully applied for accessing determinants of other matrices of moments connected to uncertainty relations, see Ref. \cite{hertz_multicopy_2019}).

We start by detailing the design of two-copy observables for accessing the determinants $d_{12}$ and $d_{14}$, used as examples to introduce the method, followed by  $d_{23}$ and $d_{15}$. Then, we increase the number of copies and consider the 3-copy nonclassicality observable for $d_{123}$ and 4-copy nonclassicality observable for $d_{1235}$. We also discuss the optical implementation of these observables whenever possible.

\subsection{Two-copy observables}

\subsubsection{Instructive examples: $d_{12}$ and $d_{14}$} \label{ssubsec:d12example}

The determinant $d_{12}$, which is expressed as 
\begin{equation}
	d_{12} =
	\begin{vmatrix}
		1 & \langle \hat{a} \rangle \\
		\langle \hatd{a} \rangle & \langle \hatd{a} \hat{a} \rangle\\
	\end{vmatrix}
	= \langle \hatd{a} \hat{a} \rangle - \langle \hat{a} \rangle \langle \hatd{a} \rangle,
\end{equation}
is useless for nonclassicality detection in the sense that it is positive for all (classical or nonclassical) states. Indeed, $d_{12}$ is simply the thermal (or chaotic) photon number, i.e., the total photon number minus the coherent photon number. Since $d_{12}$ is invariant under displacements, centering the state on the origin in phase space simply results in $d_{12} = \langle \hatd{a} \hat{a} \rangle  \ge 0$.

Nevertheless, it is instructive to illustrate the multicopy observable technique with this simple example, where we need two copies of the original state $\hat{\rho}$. We consider a $2\times 2$ operator matrix mimicking the matrix of moments $D_2$ except that we remove all expectation values. Then, we associate the first row of this operator matrix with the first copy (mode 1) and the second row with the second copy (mode 2). In a last step, we average over all permutations $\sigma \in S_2$ on the mode indices in order to ensure the Hermiticity of the resulting multicopy observable, that is,
\begin{equation}\label{eq:BrunOpB12}
\begin{split}
	\hat{B}_{12} & = \frac{1}{|S_2|}\sum_{\sigma \in S_2}
	\begin{vmatrix}
		1 &  \hat{a}_{\sigma(1)} \\
		\hatd{a}_{\sigma(2)} & \hatd{a}_{\sigma(2)} \hat{a}_{\sigma(2)} \\
	\end{vmatrix} , \\
	& = \frac{1}{2} (\hatd{a}_2 \hat{a}_2 +\hatd{a}_1 \hat{a}_1 - \hat{a}_1 \hatd{a}_2  - \hat{a}_2 \hatd{a}_1   ),
	 \end{split}
\end{equation}
where $\vert S_2 \vert = 2 !$ is the order of the symmetric group $S_2$.
The value of the determinant $d_{12}$ is obtained by measuring the expectation value of this observable on two copies of the same state $\hat{\rho}$, namely $\dmean{\hat{B}_{12}} \equiv \tr[(\hat{\rho} \otimes \hat{\rho}) \, \hat{B}_{12}] $. 
Indeed, we have
\begin{equation}
	\begin{split}
	\dmean{\hat{B}_{12}} & = \frac{1}{2} \dmean{ \, \hatd{a}_2 \hat{a}_2 +\hatd{a}_1 \hat{a}_1 - \hat{a}_1 \hatd{a}_2 - \hat{a}_2 \hatd{a}_1  \, } , \\
	& = \frac{1}{2} ( \mean{\hatd{a}_2 \hat{a}_2} + \mean{\hatd{a}_1 \hat{a}_1} - \mean{\hat{a}_1} \mean{\hatd{a}_2} - \mean{\hat{a}_2} \mean{\hatd{a}_1} ) ,\\
	& =  \langle \hatd{a} \hat{a} \rangle - \langle \hat{a} \rangle \langle \hat{a}^{\dagger} \rangle = d_{12},
	\end{split}
\end{equation}
where $\langle\cdot\rangle \equiv \tr[\hat{\rho}  \, \cdot]$.

Finally, we look for an optical implementation of observable $\hat{B}_{12}$ by means of linear optics and photon-number resolving detectors. Although it is not immediately obvious from Eq. (\ref{eq:BrunOpB12}), we can exploit the Jordan-Schwinger representation of angular momenta in terms of bosonic annihilation and creation operators (see Appendix  \ref{app:schwingerrepresentation}). The three components of the angular momentum $\hat{L}$ can be expressed as 
\begin{eqnarray}
\hat{L}_x = \frac{1}{2} (\hatd{a}_2 \hat{a}_1 + \hatd{a}_1 \hat{a}_2 ) , \\
\hat{L}_y = \frac{i}{2} (\hatd{a}_2 \hat{a}_1 - \hatd{a}_1 \hat{a}_2 ) , \label{eq-def-L_y}\\
\hat{L}_z = \frac{1}{2} (\hatd{a}_1 \hat{a}_1 - \hatd{a}_2 \hat{a}_2 ) ,
\end{eqnarray}
and they all commute with the operator
\begin{eqnarray}
\hat{L}_0 = \frac{1}{2} (\hatd{a}_1 \hat{a}_1  + \hatd{a}_2 \hat{a}_2) ,
\end{eqnarray}
with $\hat{L}_x^2+\hat{L}_y^2+\hat{L}_z^2=\hat{L}_0(\hat{L}_0+1)$ being the Casimir operator. Hence, we can express $\hat{B}_{12}$ as the difference between $\hat{L}_0$ (i.e., half the total photon number)
and the $x$-component of the angular momentum operator, namely,
\begin{equation}
	\label{equationB2}
	\hat{B}_{12} = \hat{L}_0 - \hat{L}_x.
\end{equation}
Since the application of any linear optics (passive) transformation does not change the total photon number, $\hat{L}_0$ is unaffected by such a transformation. In contrast, $\hat{L}_x$ can be turned by an appropriate linear optics transformation into $\hat{L}_z$, which corresponds to the difference of the photon numbers between the two modes. Indeed, following Ref.~\cite{hertz_multicopy_2019}, we apply a 50:50 beam splitter, which transforms the mode operators according to 
\begin{equation}\label{def:beamsplitterbalanced}
	\hat{a}'_1 = \frac{\hat{a}_1 + \hat{a}_2}{\sqrt{2}},\qquad\qquad
	\hat{a}'_2 = \frac{\hat{a}_1 - \hat{a}_2}{\sqrt{2}},
\end{equation}
so that $\hat{L}_x$ transforms into $\hat{L}_z' = \frac{1}{2} (\hat{a}'^{\dagger}_1 \hat{a}'_1  - \hat{a}'^{\dagger}_2 \hat{a}'_2)$,
where the primes refer to mode operators after the beam splitter.
Hence, the operator $\hat{B}_{12}$ is transformed into
\begin{equation}
	\hat{B}_{12} = \hat{L}_0 - \hat{L}'_z = \frac{1}{2}(\hat{n}_1'+\hat{n}_2'-\hat{n}_1'+\hat{n}_2') = \hat{n}_2'  ,
\end{equation}
where $\hat{n}_1'=\hat{a}'^{\dagger}_1 \hat{a}'_1$ and $\hat{n}_2'=\hat{a}'^{\dagger}_2 \hat{a}'_2$.

\begin{figure}[t]
	\begin{center}
		\includegraphics[width= 0.8\linewidth]{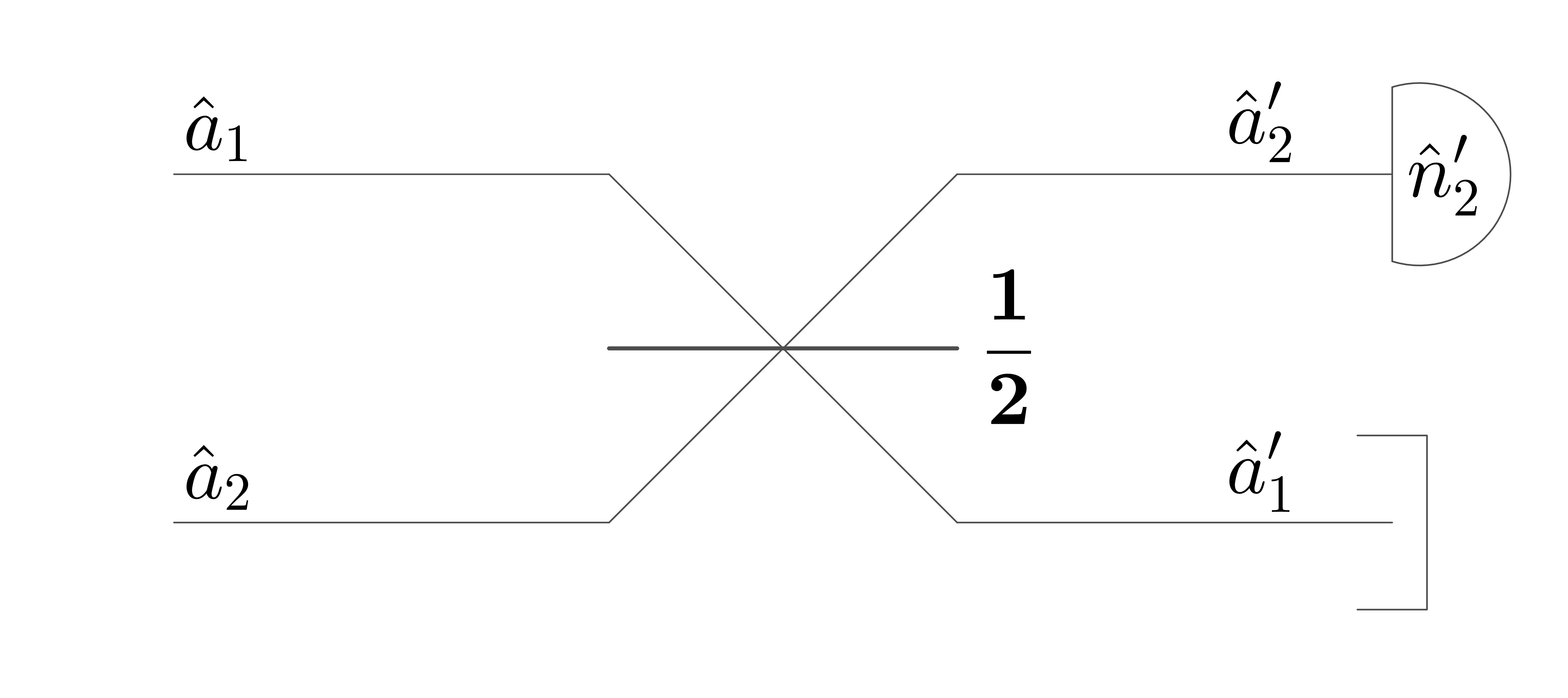}
		\caption{Implementation of the measurement of $d_{12}$. We first apply a 50:50 beam-splitter and then use a photon-number resolving detector on the second mode (i.e., the mode where the coherent fields of the two identical input states interfere destructively), yielding $n_2'$.}
		\label{Fig:Circuitd12}
	\end{center}
\end{figure}

This implies that measuring the mean photon number in the second mode after the beam splitter transformation of Fig. \ref{Fig:Circuitd12} gives the value of the determinant
\begin{equation}
	d_{12} = \dmean{\hat{B}_{12}} = \langle \hat{n}_2' \rangle.
\end{equation}
Obviously, we have $d_{12}\ge 0$, so that $d_{12}$ does not yield a useful criterion to detect nonclassical states. It is trivial to understand from Eq. (\ref{def:beamsplitterbalanced}) that this scheme gives access to the thermal (or chaotic) photon number since the coherent component of the two identical input states is concentrated on the first output mode $\hat{a}'_1$, while the mean field vanishes in the second output mode $\hat{a}'_2$. The latter is then only populated by the thermal photons.

Before moving to principal minors that are actually useful to detect nonclassicality, let us briefly consider the next case in Table \ref{table:determinantdeterminants}, namely,
\begin{equation}
	d_{14} =
	\begin{vmatrix}
		1 & \langle \hat{a}^2 \rangle \\
		\langle \hat{a}^{\dagger 2} \rangle & \langle \hat{a}^{\dagger 2} \hat{a}^2 \rangle\\
	\end{vmatrix}
	= \langle \hat{a}^{\dagger 2} \hat{a}^2 \rangle - \langle \hat{a}^2 \rangle \langle \hat{a}^{\dagger 2} \rangle .
\end{equation}
By building the corresponding two-copy observable, it is straightforward to check that $d_{14}\ge 0$ so it is useless for nonclassicality detection. Indeed, we have
	\begin{equation}
	\label{eq:BrunOpB14}
		\begin{split}
\hat{B}_{14} & = \frac{1}{|S_2|}  \sum_{\sigma \in S_2} 
	\begin{vmatrix}
			 1  &  \hatn{a}{2}_{\sigma(1)}  \\
			 \hatdn{a}{2}_{\sigma(2)}  &  \hatdn{a}{2}_{\sigma(2)} \hatn{a}{2}_{\sigma(2)}  \\
		\end{vmatrix} ,\\
		& = \frac{1}{2}(\hatdn{a}{2}_2
		 \hatn{a}{2}_2 + \hatdn{a}{2}_1
		 \hatn{a}{2}_1 - \hatn{a}{2}_1 \hatdn{a}{2}_2
		  - \hatn{a}{2}_2 \hatdn{a}{2}_1
		 ) ,
		\end{split}
	\end{equation}
In analogy with $\hat{B}_{12}$, this two-copy observable can be reexpressed in terms of angular momentum operators, namely,
\begin{equation}
	\hat{B}_{14} = 2 \, (\hat{L}_0^2 - \hat{L}_x^2) .
\end{equation}
As before, we may transform $\hat{L}_x$ into $\hat{L}_z'$ by using a 50:50 beam splitter as described in Eq. (\ref{def:beamsplitterbalanced}), which 
gives
\begin{equation}
\hat{B}_{14} = 2 \, (\hat{L}_0^2 - \hat{L}_z^{'2})
=  2 \, \hat{n}_1' \hat{n}_2'  .
\end{equation}
Thus, this determinant can be accessed by applying a 50:50 beam splitter on two identical copies as in Fig.~\ref{Fig:Circuitd12} but then measuring the mean value of the product of the photon numbers, that is
\begin{equation}
\label{eq:d14>=0}
	d_{14} = \dmean{\hat{B}_{14}} = 2 \, \langle \hat{n}_1' \hat{n}_2' \rangle   \ge 0.
\end{equation}

In the following, we apply the same technique to determinants that enable the detection of nonclassicality. The calculations follow exactly the same path: we assign a mode to each row of the operator matrix and then symmetrize it as in Eq. (\ref{eq:BrunOpB12}) or (\ref{eq:BrunOpB14}). Finally, whenever possible, we find a linear optics transformation such that the observable can be measured by means of photon number resolving detectors. Since the difficulty of this procedure increases with the number of copies, we limit our search to principal submatrices of $D_5$ up to dimension $4 \times 4$.

\subsubsection{Detection of squeezed states: $d_{23}$}

As shown in Table \ref{table:determinantdeterminants}, the two most interesting principal submatrices of dimension $2 \times 2$ for detecting nonclassical states are $d_{15}$ and $d_{23}$. We start with $d_{23}$, expressed as 
	\begin{equation}\label{def:d23criteria}
		d_{23} = 
		\begin{vmatrix}
			\langle \hatd{a} \hat{a} \rangle & \langle \hatdn{a}{2} \rangle \\
			\langle \hat{a}^2 \rangle & \langle \hatd{a} \hat{a} \rangle\\
		\end{vmatrix}
		= \langle \hatd{a} \hat{a} \rangle^2 - \langle \hat{a}^2 \rangle \langle \hat{a}^{\dagger 2} \rangle.
	\end{equation}
	The criterion derived from $d_{23}$ detects squeezed states and even cat states (it does not detect Fock states and odd cat states). Note that $d_{23}$ is not invariant under displacements (as we shall see, this invariance can be enforced by considering $d_{123}$ instead). Following the procedure described in Sec.~\ref{ssubsec:d12example}, we obtain the multicopy observable
	\begin{equation}
		\begin{split}
\hat{B}_{23} & = \frac{1}{|S_2|}  \sum_{\sigma \in S_2} 
	\begin{vmatrix}
			 \hatd{a}_{\sigma(1)} \hat{a}_{\sigma(1)}  &  \hatdn{a}{2}_{\sigma(1)}  \\
			 \hat{a}^2_{\sigma(2)}  &  \hatd{a}_{\sigma(2)} \hat{a}_{\sigma(2)}  \\
		\end{vmatrix} ,\\
		& = \hatd{a}_1 \hat{a}_1 \hatd{a}_2 \hat{a}_2 - \frac{1}{2} \left( \hatdn{a}{2}_1  \hat{a}^2_2  + \hatdn{a}{2}_2  \hat{a}^2_1   \right).
		\end{split}
	\end{equation}
	Similarly as for $\hat{B}_{12}$ or $\hat{B}_{14}$, we can express $\hat{B}_{23}$ in terms of angular momentum operators, namely,
	\begin{align}
	\label{eq:B_23operator}
	\begin{split}
		\hat{B}_{23} = 2 \, \hat{L}_y^2- \hat{L}_0 .		
		\end{split}
	\end{align}
It must be noted that Eq.	(\ref{eq:B_23operator}) can also be reexpressed more concisely as a normally ordered operator, namely,
\begin{equation}
	\hat{B}_{23} = 2 \, : \! \hat{L}_y^2 \! : ~,
	\label{eq-B23-normally-ordered}
\end{equation}
where the normal ordering symbol must be understood term by term, that is, we must expand $\hat{L}_y^2$ in powers of $\hat{a}$ and $\hatd{a}$ and then normally order each term separately.

\begin{figure}[t]
		\begin{center}
			\includegraphics[width=\linewidth]{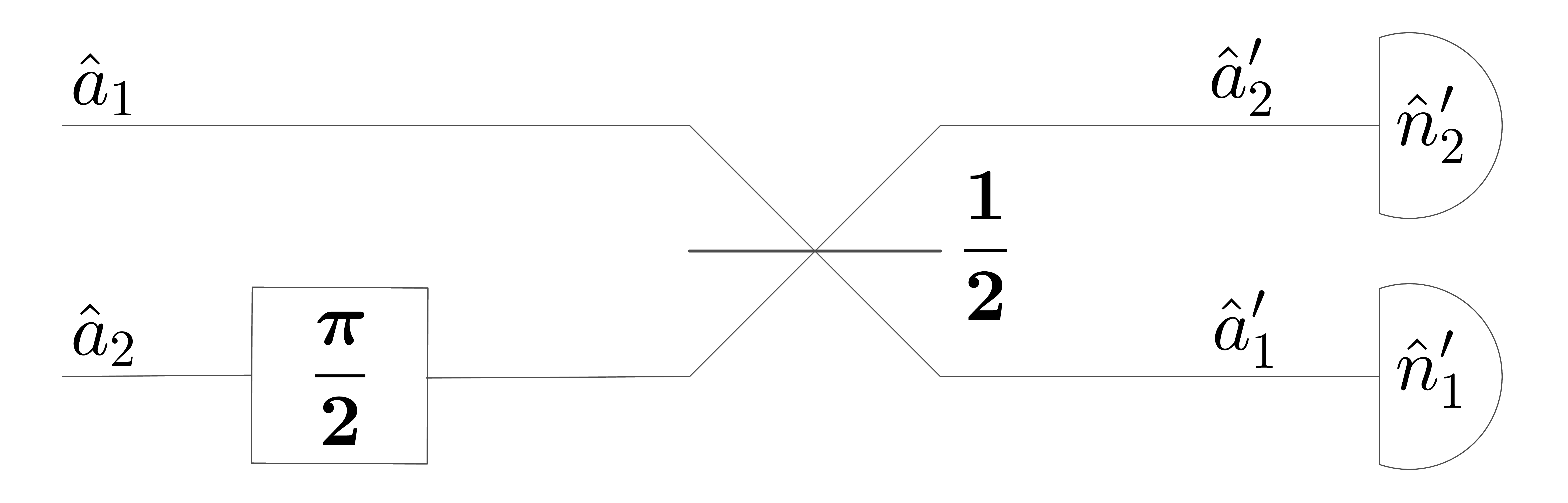}
			\caption{Implementation of the measurement of $d_{23}$. We first apply a phase shift of phase $\pi/2$ on the second mode, followed by a 50:50 beam splitter. Then, the value of $d_{23}$ is accessed by measuring the number of photons in both output modes and computing Eq. (\ref{eq:d23observable}).}
			\label{Circuit_d23}
		\end{center}
	\end{figure}

Using Eq. (\ref{eq:B_23operator}), we may design a linear interferometer in order to measure $\hat{B}_{23}$ with photon-number resolving detectors. We consider the same interferometer as considered in Ref. \cite{hertz_multicopy_2019}, which is composed of a $\pi/2$ phase shifter on the second mode followed by a 50:50 beam splitter as shown in Fig. \ref{Circuit_d23}. Under the $\pi/2$ local phase shift, the second mode operator transforms according to 
\begin{equation}\label{def:phaseshift}
\hat{a}'_2 = - i \, \hat{a}_2,
\end{equation}
while the 50:50 beam splitter transformation is described in Eq. (\ref{def:beamsplitterbalanced}). The operator $\hat{L}_0$ is again invariant under these operations, but $\hat{L}_y$ transforms into $\hat{L}_x'$ following the phase shifter on the second mode and then transforms into $\hat{L}_z'$ after the 50:50 beam splitter.
Hence, after applying the interferometer shown in Fig. \ref{Circuit_d23}, the nonclassicality observable takes the form
\begin{equation}
\label{eq:d23observableB23}
		\begin{split}
		\hat{B}_{23} &= 2 \, : \! \hat{L}_z^{' 2} \! : ~,\\
		&= 2 \, \hat{L}_z^{' 2} - \hat{L}_0 ,\\
		&=  \frac{1}{2}(\hat{n}_1' - \hat{n}_2')^2 - \frac{1}{2}(\hat{n}_1' + \hat{n}_2')  ,
	    \end{split}
	\end{equation}
and its expectation value yields
	\begin{equation}
	\label{eq:d23observable}
		\begin{split}
		d_{23} &= \dmean{\hat{B}_{23}}  ,\\
		&= \frac{1}{2} \left\langle (\hat{n}_1' - \hat{n}_2')^2 - (\hat{n}_1' + \hat{n}_2') \right\rangle .
	    \end{split}
	\end{equation}
As a consequence, the principal minor $d_{23}$ can be evaluated simply by accessing the joint photon number statistics on the two output modes $\hat{a}_1'$ and $\hat{a}_2'$.

It is instructive to understand how the nonclassicality of a squeezed state is detected by Eq. (\ref{eq:d23observable}). Two copies of a squeezed state are transformed through the interferometer of Fig. \ref{Circuit_d23} as follows. The phase shift rotates the second squeezed states by $\pi/2$, and the 50:50 beam splitter produces (from the two orthogonal squeezed states) a two-mode squeezed vacuum state, $(\cosh r)^{-1} \sum_{n=0}^{\infty} (\tanh r)^n ~| n,n \rangle$. This state exhibits a perfect photon-number correlation. Hence, the squared photon-number difference in Eq. (\ref{eq:d23observable}) vanishes while the second term, which is proportional to the sum of photon numbers, comes with a negative sign. Thus, squeezed states are detected as nonclassical with $d_{23}<0$ as soon as $r>0$.

Note that Eq. (\ref{eq:d23observable}) can also be reformulated as

\begin{equation}
\label{eq_d23_Mandel}
	d_{23} = \frac{1}{2}\left(Q'_1 \langle \hat{n}'_1 \rangle + Q'_2 \langle \hat{n}'_2 \rangle + \langle \hat{n}'_1 \rangle^2 + \langle \hat{n}'_2 \rangle^2 - 2 \langle \hat{n}'_1 \hat{n}'_2\rangle\right), 
\end{equation}
where $Q'_1$ and $Q'_2$ denote the Mandel $Q$ parameters of the output states occupying modes $\hat{a}'_1$ and $\hat{a}'_2$. The $Q$ parameter is defined as
\begin{equation}
	Q = \frac{(\Delta \hat{n})^2 - \mean{\hat{n}}}{\mean{\hat{n}}},
\end{equation}
and measures the deviation from ``Poissonianity'' of the state (it vanishes for a coherent state, associated with a Poisson distribution).
If the input state is a product of two identical coherent states, it is transformed under the interferometer of Fig. \ref{Circuit_d23} into a product of two coherent states, hence $Q'_1=Q'_2=0$. Further,  $\langle \hat{n}'_1 \rangle^2 = \langle \hat{n}'_2 \rangle^2 = \langle \hat{n}'_1 \hat{n}'_2\rangle$ since the two output coherent states are independent and have equal squared amplitudes. This confirms that $d_{23} =0$ for coherent states.

\subsubsection{Detection of Fock states: $d_{15}$}
	
The criterion based on $d_{15}$ is complementary to the one based on $d_{23}$ as it detects Fock states and odd cat states (it does not detect squeezed states and even cat states). It is defined as 	
	\begin{equation}\label{def:d15criteria}
		d_{15} =
		\begin{vmatrix}
			1 & \langle \hatd{a} \hat{a} \rangle \\
			\langle \hatd{a} \hat{a} \rangle & \langle \hat{a}^{\dagger 2} \hat{a}^2 \rangle\\
		\end{vmatrix}
		= \mean{\hat{a}^{\dagger 2} \hat{a}^2} - \mean{\hatd{a} \hat{a}}^2,
	\end{equation}
and is not invariant under displacements (just as $d_{23}$).
It can be rewritten as
\begin{equation}
	d_{15} = \mean{\hat{n}^2} - \mean{\hat{n}}^2 - \mean{\hat{n}} = (\Delta \hat{n})^2 - \mean{\hat{n}},
	\end{equation}
and can thus be reexpressed in terms of the Mandel $Q$ parameter of the input state as	\begin{equation}
\label{eq_d15_Mandel}
	d_{15} = Q \, \mean{\hat{n}} ,
\end{equation}
so that the nonclassicality criterion based on $d_{15}$ is simply a witness of the sub-Poissonian statistics ($Q<0$) of the state. Obviously, we have $d_{15}=0$ for coherent states while $d_{15}>0$ for (classical) thermal states, as expected. 

The procedure described in Sec.~\ref{ssubsec:d12example} yields the following two-copy nonclassicality observable 
	\begin{equation}
		\hat{B}_{15} =  \frac{1}{2}(\hat{n}_1- \hat{n}_2)^2 - \frac{1}{2}(\hat{n}_1 + \hat{n}_2) ,
	\end{equation}
whose expectation value is written as
	\begin{equation}\label{eq:d15numberop}
		d_{15} = \frac{1}{2} \left\langle (\hat{n}_1- \hat{n}_2)^2 - (\hat{n}_1 + \hat{n}_2) \right\rangle.
	\end{equation}
Interestingly, $d_{15}$ involves the same observable as the one used to measure $d_{23}$ [see Eq.~(\ref{eq:d23observable})] except that we do not need the prior interferometer. This similarity will be exploited in the next Section. It also implies that $d_{15}$ can be expressed in terms of Mandel $Q$ parameters of the input states, namely
\begin{equation}
	d_{15} = \frac{1}{2}\left(Q_1 \langle \hat{n}_1 \rangle + Q_2 \langle \hat{n}_2 \rangle + (\langle \hat{n}_1 \rangle - \langle \hat{n}_2 \rangle)^2 \right), 
\end{equation}
which resembles Eq. \eqref{eq_d23_Mandel} where we have used 	
$\langle \hat{n}_1 \hat{n}_2\rangle=\langle \hat{n}_1 \rangle \, \langle \hat{n}_2 \rangle$ since the inputs are in a product state. Of course, for two identical inputs, this reduces to Eq.  \eqref{eq_d15_Mandel}.

\subsubsection{Interpolation between $d_{15}$ and $d_{23}$} \label{ssubsec:twomodeinterpolation}

\begin{figure}[t]
	\begin{center}
		\includegraphics[width=0.9\linewidth]{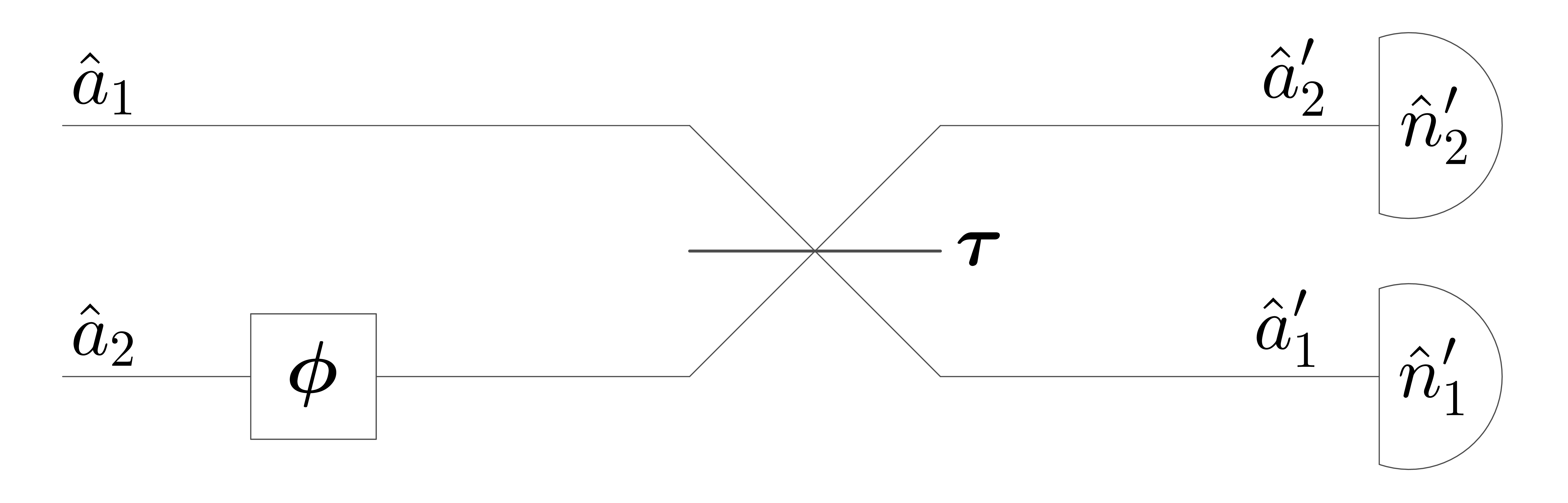}
		\caption{Extended circuit that interpolates between the measurement of $d_{15}$ and $d_{23}$. We need  a phase shifter of phase $\phi$ followed by a beam splitter of transmittance $\tau$, and then we measure the photon number on both output modes in order to access the expectation value of the observable of Eq. (\ref{eq:B1523out}). }
		\label{fig:circuit2modes}
	\end{center}
\end{figure}

As we observe in Table \ref{table:determinantdeterminants}, the criteria $d_{15}$ and $d_{23}$ taken together detect the four considered kinds of pure states. Given the similarity between 
Eqs.~(\ref{eq:d23observable}) and (\ref{eq:d15numberop}), it is tempting to construct a common multicopy observable that interpolates between $d_{15}$ and $d_{23}$. It is based on a linear optical interferometer composed of a phase shifter of phase $\phi$ and a beam splitter of transmittance $\tau$ (see Fig. \ref{fig:circuit2modes}), followed by the measurement of the observable 
\begin{equation}\label{eq:B1523out}
\hat{B}_{15,23}= \frac{1}{2}(\hat{n}_1'- \hat{n}_2')^2 - \frac{1}{2}(\hat{n}_1' + \hat{n}_2') ,
\end{equation} 
where the primes refer to output modes.
By applying the interferometer of Fig. \ref{fig:circuit2modes} backwards on Eq. (\ref{eq:B1523out}), we may reexpress it as a function of the input mode operators, namely
\begin{align}\label{eq:B1523in}
\begin{split}
	\hat{B}_{15,23} = & \frac{1}{2}:[(\hatd{a}_2 \hat{a}_1 \, e^{- i \phi} + \hatd{a}_1 \hat{a}_2 \, e^{i \phi} ) \, 2 \sqrt{(1 - \tau) \tau} \\
	& + (\hatd{a}_1 \hat{a}_1  - \hatd{a}_2 \hat{a}_2 ) \, (-1 + 2 \tau)]^2:.
	\end{split}
\end{align}
This observable clearly interpolates between $\hat{B}_{15}$ ($\tau = 1$) and $\hat{B}_{23}$ ($\phi = \pi /2$ and $\tau = 1/2$). Since Eq. (\ref{eq:B1523in}) is the square of a Hermitian operator, $\hat{B}_{15,23}$ can be written under the form $:\!\! \hat{f}^\dagger \hat{f}\!\! :$
 and is indeed a valid observable for witnessing nonclassicality. 

\begin{figure}[t!]
	\begin{center}
		\includegraphics[width=0.99\linewidth, height=5cm]{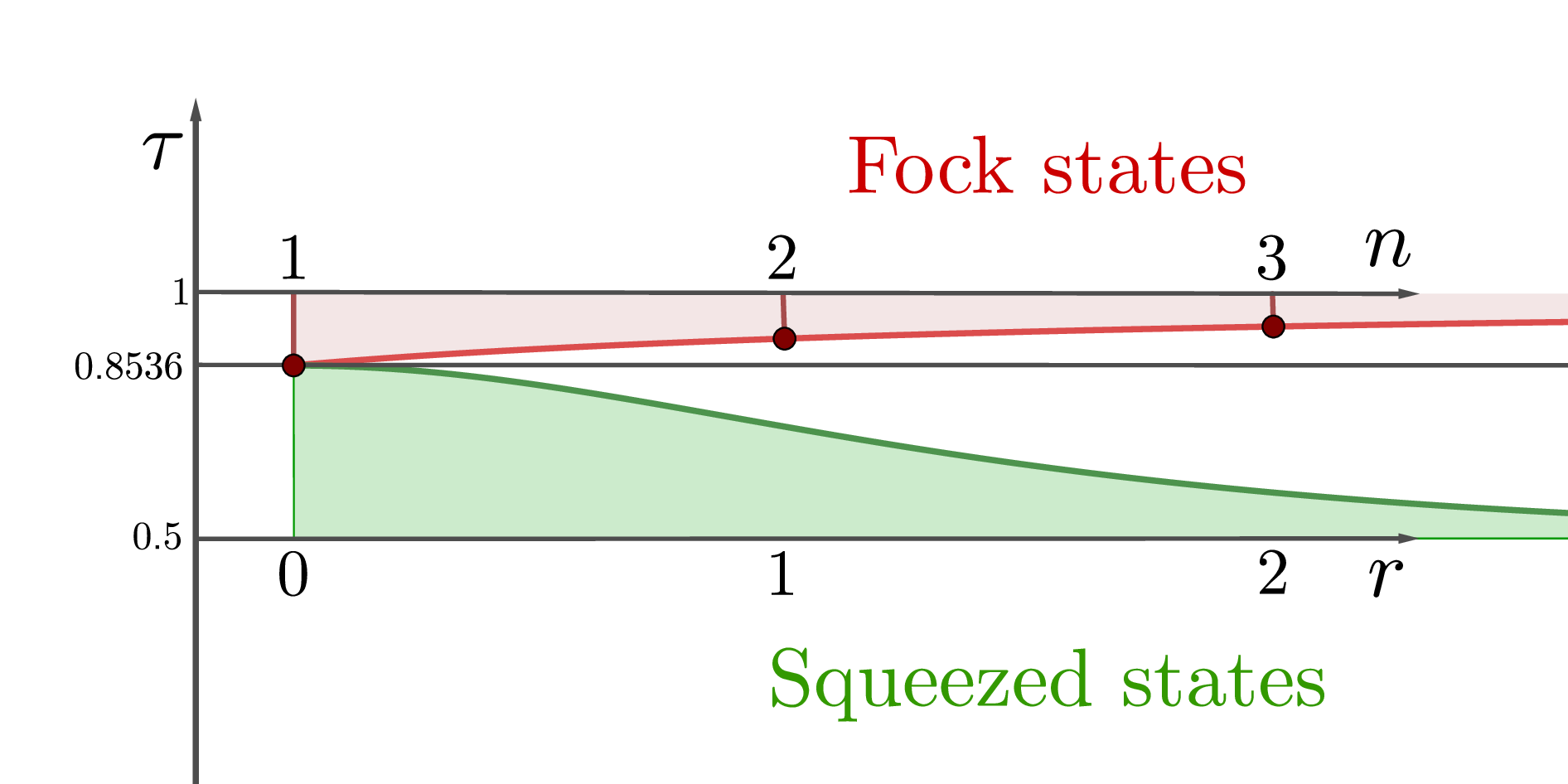}
		\caption{Detection limit on the parameter ($n$ or $r$) characterizing the input state for several values of the transmittance $\tau$ (the phase is set to $\phi=\pi/2$). The red area corresponds to the values of $n$ where Fock states are detected (all of them are detected for $\tau=1$, reducing to $d_{15}$), while the green region corresponds to the values of $r$ where squeezed states are detected (all of them are detected for $\tau=1/2$, reducing to $d_{23}$). Unfortunately, at the threshold value $\tau^* \approx 0.8536$, all Fock states and squeezed states are left undetected.}
		\label{fig:d1523detection}
	\end{center}
\end{figure}

We start by setting the phase shift to $\phi = \pi /2$ as it does not play any role in $\hat{B}_{15}$ and study the detection of the different types of nonclassical states as a function of the transmittance $\tau$ (with $1/2\le \tau \le 1$), see Fig.~\ref{fig:d1523detection}. 
It appears that Fock states $\ket{n}$ with $n\ge 1$ may only be detected when the transmittance is above a threshold value $\tau^* = (2+\sqrt{2})/4 \approx 0.8536$, while squeezed states with $r>0$ may only be detected for values of the transmittance lower than this threshold $\tau^*$. The value of $\tau^*$ is such that the coefficients of the two operator terms in Eq. (\ref{eq:B1523in}) are equal. Moreover, it is possible to show that the odd cat states may only be detected when $\tau>\tau^*$ while the even cat states may only be detected when $\tau<\tau^*$. This means that there is no value of the transmittance enabling the detection of the four classes of nonclassical states considered in Table \ref{table:determinantdeterminants}. Unfortunately, changing the value of the phase shift $\phi$ does not change the situation, so we cannot find a single two-copy observable that detects all four classes of nonclassical states.

Note that the criteria based on $d_{15}$ and $d_{23}$ can be viewed as complementary: if one of them detects a nonclassical state, i.e., its value is negative, then the other one is necessarily positive for that state (of course, they can be both positive as, for example, for classical states). Indeed, we have

\begin{equation}\label{eq:complementarityd15d23}
 d_{23} + d_{15} = d_{14}  \geq 0,
\end{equation}
where the inequality comes from Eq. (\ref{eq:d14>=0}). Hence, the witnesses $d_{15}$ and $d_{23}$ cannot both simultaneously detect nonclassicality for a given state, as illustrated for a superposition of three Fock states in Appendix \ref{app:complementarityexample}.

\subsubsection{Effect of a displacement on $d_{15}$ and $d_{23}$} \label{ssubsec:twomodeinterpolation+displacement}

In general, we expect that the nonclassical character of a quantum state will be harder to detect when the state is moved away from the origin in phase space. As we shall see, this is often (but not always) the case. We can calculate the difference $\Delta^{\alpha}$ between the determinant ($d_{15}$ or $d_{23}$) when the state is displaced by $\hat{D}(\alpha)$ with $\alpha = |\alpha| e^{i \theta_\alpha}$ and the same determinant when the state is centered. The differences $\Delta^{\alpha}$ in the case of $d_{15}$ are presented in Table \ref{table:effectdisplacement} for the considered states.

For Fock states as well as odd cat states, the effect of a displacement on $d_{15}$ is given by an extra positive factor $\Delta^{\alpha}>0$, so that displacements always deteriorate the detection.
For squeezed states as well as even cat states, the result of a displacement on $d_{15}$ is that it can either enhance ($\Delta^{\alpha} < 0$) or deteriorate ($\Delta^{\alpha} > 0$) the detection of nonclassicality. Indeed, the sign of $\Delta^{\alpha}$ depends on the difference between the angle of squeezing $\psi$ (or the angle of the cat state $\theta_{\beta}$) and the angle of the displacement $\theta_{\alpha}$.

\subsection{Three-copy observable}

As we have observed in Sec. \ref{ssubsec:twomodeinterpolation+displacement}, the two-mode criteria $d_{23}$ and $d_{15}$ are not invariant under displacements. This comes with the fact that some nonclassical states become undetected if they are displaced in phase space. In order to overcome this effect of displacements, we build an observable involving a third replica of the input state following a similar reasoning as in Ref. \cite{hertz_multicopy_2019}. We focus on the three-copy observable $\hat{B}_{123}$, which can be obtained by extending the procedure described in Sec.~\ref{ssubsec:d12example}. From the explicit form of the principal minor
\begin{align}
\begin{split}
	d_{123}  = ~ &
	\begin{vmatrix}
		1 & \mean{\hat{a}} & \mean{\hatd{a}} \\
		\mean{\hatd{a}} & \mean{\hatd{a} \hat{a}} & \mean{\hat{a}^{\dagger 2}} \\
		\mean{\hat{a}} & \mean{\hat{a}^2} & \mean{\hatd{a} \hat{a}} \\
	\end{vmatrix}, \\
	 = ~ & \mean{\hatd{a} \hat{a}}^2 - \mean{\hat{a}^{\dagger 2}} \mean{\hat{a}^2} 
	- 2 \mean{\hatd{a}} \mean{\hat{a}} \mean{\hatd{a} \hat{a}} \\
	& + \mean{\hat{a}^{\dagger 2}} \mean{\hat{a}}^2 
	+ \mean{\hatd{a}}^2 \mean{\hat{a}^2},
	\end{split}
\end{align}
we get the corresponding nonclassicality observable
\begin{align}
\begin{split}  \label{eq:B123explicit}
	\hat{B}_{123} & = \frac{1}{|S_3|}  \sum_{\sigma \in S_3}
	\begin{vmatrix}
		1 &  \hat{a}_{\sigma(1)} &  \hatd{a}_{\sigma(1)} \\
		 \hatd{a}_{\sigma(2)} &  \hatd{a}_{\sigma(2)} \hat{a}_{\sigma(2)} &  \hat{a}^{\dagger 2}_{\sigma(2)} \\
		 \hat{a}_{\sigma(3)} &  \hat{a}^2_{\sigma(3)} &  \hatd{a}_{\sigma(3)} \hat{a}_{\sigma(3)} \\
	\end{vmatrix}, \\
	& = \frac{1}{3} \left(\hatd{a}_2 \hat{a}_2   \hatd{a}_3 \hat{a}_3 +  \hatd{a}_1 \hat{a}_1  \hatd{a}_3 \hat{a}_3 + \hatd{a}_1 \hat{a}_1   \hatd{a}_2 \hat{a}_2 \right) \\
	& - \frac{1}{6} (\hatdn{a}{2}_2   \hat{a}^2_3 + \hat{a}^2 _2   \hatdn{a}{2}_3 +  \hatdn{a}{2}_1 \hat{a}^2_3 + \hatn{a}{2}_1 \hatdn{a}{2}_3 + \hatdn{a}{2}_1 \hat{a}^2_2 + \hat{a}^2_1 \hatdn{a}{2}_2   ) \\
	& - \frac{1}{3} (\hatd{a}_1 \hat{a}_1   \hatd{a}_2   \hat{a}_3 + \hatd{a}_1 \hat{a}_1   \hat{a}_2   \hatd{a}_3 + \hat{a}_1   \hatd{a}_2 \hat{a}_2   \hatd{a}_3 \\
	& ~~~~~ + \hatd{a}_1   \hatd{a}_2 \hat{a}_2   \hat{a}_3 + \hat{a}_1   \hatd{a}_2   \hatd{a}_3 \hat{a}_3 + \hatd{a}_1   \hat{a}_2   \hatd{a}_3 \hat{a}_3 ) \\
	& + \frac{1}{3} \left( \hatdn{a}{2}_1   \hat{a}_2    \hat{a}_3 + \hat{a}_1   \hatdn{a}{2}_2    \hat{a}_3  + \hat{a}_1    \hat{a}_2   \hatdn{a}{2}_3  \right) \\
	& + \frac{1}{3} \left( \hat{a}^2_1    \hatd{a}_2   \hatd{a}_3 + \hatd{a}_1   \hat{a}^2_2   \hatd{a}_3 + \hatd{a}_1   \hatd{a}_2   \hat{a}^2_3 \right) , 
\end{split}
\end{align}
where $\vert S_3 \vert = 3 !$. It is straightforward to check  that the mean value of this observable (\ref{eq:B123explicit}) over three identical copies gives $\dmean{\hat{B}_{123}}=d_{123}$.

\small
\begin{table}
\centering
$ 
\begin{array}{|c|c|}
\hline

\text{State}  &  \Delta^{\alpha} = d_{15}^{\alpha} - d_{15}  \\ 
\hline
\text{Fock} & \Delta^{\alpha} = 2 n | \alpha |^2 \\
\text{Squeezed} & \Delta^{\alpha} = 2 \sinh{(r)} | \alpha |^2 (\cosh{(r)} \cos{(2 \theta_\alpha - \psi)} + \sinh{(r)}) \\
\text{Odd cat} & \Delta^{\alpha} = 2 | \alpha |^2 | \beta |^2 ( \cos{(2 \theta_\alpha - 2 \theta_\beta)}+ \frac{N_+}{N_-}) \\ 
\text{Even cat} & \Delta^{\alpha} = 2 | \alpha |^2 | \beta |^2 (\frac{N_-}{N_+} - \cos{(2 \theta_\alpha - 2 \theta_\beta)}) \\
\hline
\end{array}
$ 
\caption{Effect of the displacement $\hat{D}(\alpha)$ on the determinant $d_{15}$. The table shows the difference between the determinant when the state is displaced and when it is centered (so that $\Delta^{\alpha} < 0$ implies an enhanced detection capability).}
\label{table:effectdisplacement}
\end{table}
\normalsize

\begin{figure}[t]
	\begin{center}
		\includegraphics[width=1 \linewidth]{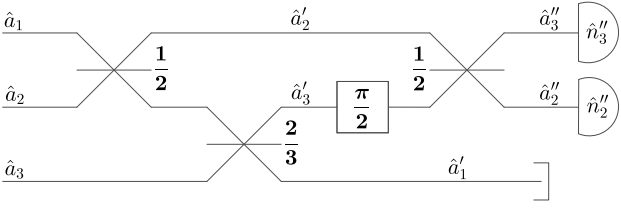}
		\caption{Three-mode circuit for accessing the  nonclassicality witness $d_{123}$. We recognize the circuit for measuring $d_{23}$ applied on modes 2 and 3, preceded by two beam splitters of transmittance $1/2$ and $2/3$ corresponding to the transformation of Eq. (\ref{eq-rotation-a-ad}). The role of these two beam splitters is to concentrate the coherent field of the three replicas into mode 1, which is left unmeasured.}
		\label{fig:circuit3modes}
	\end{center}
\end{figure}

Similarly to $\hat{B}_{23}$, the nonclassicality observable $\hat{B}_{123}$ can be written in a much more compact form in terms of a normally-ordered expression 
\begin{equation}
\label{eq_B123_normally_ordered}
\hat{B}_{123} =~  \frac{2}{3} : \!  \left( \hat{L}^{12}_y + \hat{L}^{23}_y + \hat{L}^{31}_y  \right)^2 \! : ~ ,
\end{equation}
where $\hat{L}^{kl}_y=\frac{i}{2} (\hatd{a}_l \hat{a}_k - \hatd{a}_k \hat{a}_l)$. Here, the superscript of the angular momentum component $\hat{L}_y$ stands for the two modes that are involved in the definition (\ref{eq-def-L_y}). The observable $\hat{B}_{123}$ can be accessed by first applying a linear optics transformation corresponding to the first two beam splitters in Fig.~\ref{fig:circuit3modes}, which effects the rotation 
\begin{equation}
\label{eq-rotation-a-ad}
\begin{split}
\hat{a}'_1 &=\frac{1}{\sqrt{3}}(\hat{a}_1+\hat{a}_2+\hat{a}_3) , \\
\hat{a}'_2 &= \frac{1}{\sqrt{2}}(\hat{a}_1-\hat{a}_2) , \\
\hat{a}'_3 &=\frac{1}{\sqrt{6}}(\hat{a}_1+\hat{a}_2 - 2 \hat{a}_3) ,
\end{split}
\end{equation}
on the mode operators. Interestingly, this induces the same rotation of the angular momentum $y$\mbox{-}components,
\begin{equation}
\label{eq-rotation-L_y}
\begin{split}
\hat{L}_y^{23 '} &=\frac{1}{\sqrt{3}}(\hat{L}_y^{23}+\hat{L}_y^{31}+\hat{L}_y^{12}) , \\
\hat{L}_y^{31 '} &= \frac{1}{\sqrt{2}}(\hat{L}_y^{23}-\hat{L}_y^{31}) ,\\
\hat{L}_y^{12 '} &=\frac{1}{\sqrt{6}}(\hat{L}_y^{23}+\hat{L}_y^{31} - 2 \hat{L}_y^{12}) .
\end{split}
\end{equation}
In other words, the two vectors $(\hat{a}_1,\hat{a}_2 ,\hat{a}_3)^T$ and $(\hat{L}_y^{23},\hat{L}_y^{31},\hat{L}_y^{12})^T$ undergo the exact same orthogonal transformation. Note that, in contrast, the vectors $(\hat{L}_x^{23},\hat{L}_x^{31},\hat{L}_x^{12})^T$ and $(\hat{L}_z^{23},\hat{L}_z^{31},\hat{L}_z^{12})^T$ undergo different linear transformations which are mixing their components. 
By using Eq. (\ref{eq-rotation-L_y}), we can reexpress Eq.~(\ref{eq_B123_normally_ordered}) in terms of output angular momentum variables as
\begin{equation}
\label{eq:B123operator_1st_expression}
\hat{B}_{123} =~  2 : \!  \left( \hat{L}_y^{23'} \right)^2 \! :  ~ ,
\end{equation}
which resembles Eq. (\ref{eq-B23-normally-ordered}) except that it acts on modes 2 and 3. Hence, in analogy with what we did for $\hat{B}_{23}$, we can access $\hat{B}_{123}$ with a subsequent linear optical transformation applied onto modes 2 and 3, corresponding to the phase shifter and last beam splitter in Fig.~\ref{fig:circuit3modes}. This converts the angular momentum $y$-component associated with modes 2 and 3 into the corresponding $z$\mbox{-}component, so we have 
\begin{equation}
		\begin{split}
		\hat{B}_{123} &= 2  : \! \left( \hat{L}_z^{23 ''} \right)^2 \!\! : ~,\\
		&= 2 \, \left( \hat{L}_z^{23 ''} \right)^2 - \hat{L}_0^{23} , 
	    \end{split}
	\end{equation}
where the double primes refer to the output of the full circuit of Fig.~\ref{fig:circuit3modes}. Here $\hat{L}_z^{kl}=\frac{1}{2} (\hatd{a}_k \hat{a}_k - \hatd{a}_l \hat{a}_l)$ and $\hat{L}_0^{kl}=\frac{1}{2} (\hatd{a}_k \hat{a}_k + \hatd{a}_l \hat{a}_l)$.
Thus, after applying this circuit, the observable $\hat{B}_{123}$ transforms into
\begin{equation}\label{eq:B123operator}
\hat{B}_{123} = \frac{1}{2}(\hat{n}''_2-\hat{n}''_3)^2-\frac{1}{2}(\hat{n}''_2+ \hat{n}''_3)  ,
\end{equation}
which is analogous to Eq. (\ref{eq:d23observableB23}). Its expectation value yields
	\begin{equation}
		\begin{split}
		d_{123} &= \dmean{\hat{B}_{123}},  \\
		&= \frac{1}{2} \left\langle (\hat{n}''_2-\hat{n}''_3)^2 - (\hat{n}''_2+ \hat{n}''_3) \right\rangle .
	    \end{split}
	\end{equation}
The intuition behind this circuit follows from Ref. \cite{hertz_multicopy_2019}. With the first two beam splitters in Fig.~\ref{fig:circuit3modes}, we apply a transformation that concentrates the coherent component of the three identical input states in the first mode, which is traced out. The second and third modes thus have a vanishing mean field, so we can apply the same scheme as for measuring $\hat{B}_{23}$ but on these two modes, which leads to $\hat{B}_{123}$.
Up to a mode relabelling, the operator in Eq. (\ref{eq:B123operator}) is indeed the same as the one used for evaluating $d_{23}$, see Eq.~(\ref{eq:d23observable}). Notably, the circuit for evaluating $d_{123}$ as shown in Fig.~\ref{fig:circuit3modes} is the same circuit as the one used in Ref. \cite{hertz_multicopy_2019} in order to measure an uncertainty observable; the key difference is that the observable  $(\hat{n}''_2-\hat{n}''_3)^2 /4$ must be measured instead at the output of the circuit in order to evaluate the uncertainty.

It must be noted that the circuit of Fig.~\ref{fig:circuit3modes} is not unique. Another option to access $\hat{B}_{123}$ is to apply the circuit of a three-dimensional discrete Fourier transform on the three input modes. This circuit is actually equivalent to the circuit of Fig.~\ref{fig:circuit3modes} up to $\pm \pi/6$ phase shifters on modes 2 and 3, which do not play a role since we measure photon numbers. Hence, this leads to the same observable.

\begin{figure}[t!]
	\begin{center}
		\includegraphics[width=1 \linewidth]{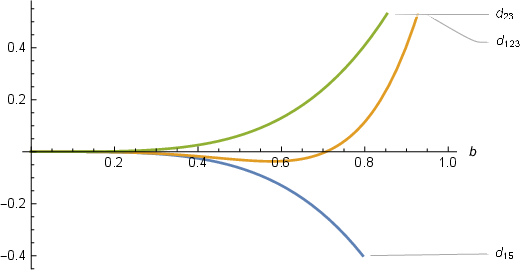}
		\caption{Comparison of the values of $d_{123}$ and $d_{23}$ for a superposition of $\vert 0 \rangle$  and $\vert 1 \rangle$ Fock states. While $d_{23}$ never detects any such superposition as nonclassical, $d_{123}$ detects them up to $b=0.7$. Moreover, $d_{15}$ always detects these superposition states as nonclassical.}
		\label{fig:d123VSd23}
	\end{center}
\end{figure}

We stress that in addition of being the displacement-invariant version of $d_{23}$, the criterion based on $d_{123}$ is also superior in that it is able to detect states that are different from displaced states detected by $d_{23}$. A simple example is the superposition of the two Fock states $\vert 0 \rangle$  and $\vert 1 \rangle$ as given in Appendix \ref{app:complementarityexample}, see Eq. (\ref{def:012superposition}) with $c=0$. The values of $d_{23}$ and $d_{123}$ are plotted in Fig. \ref{fig:d123VSd23}. We see that while $d_{23}$ never detects any superposition of the first two Fock states, $d_{123}$ detects such superpositions up to $b=0.7$. Note that $d_{15}$ is always negative, which confirms that these superpositions are always nonclassical.

\subsection{Four-copy observable}

By adding a fourth replica of the input state, it is possible to further improve the detection capability of nonclassicality observables. The most interesting nonclassicality criterion derived from a $4 \times 4$ matrix is $d_{1235}$ since it detects the nonclassicality of all squeezed, Fock, and even or odd cat states (see Table \ref{table:determinantdeterminants}). It is written as
\begin{equation}
	d_{1235} = \begin{vmatrix}
		1 & \mean{\hat{a}} & \mean{\hatd{a}} &     \langle \hat{a}^\dagger \hat{a} \rangle \\
		\mean{\hatd{a}}  &    \langle \hatd{a} \hat{a} \rangle&  \langle   \hatdn{a}{2} \rangle & \langle \hat{a}^{\dagger 2} \hat{a} \rangle\\
		\mean{\hat{a}} &     \langle \hat{a}^2 \rangle & \langle     \hatd{a} \hat{a} \rangle & \langle     \hatd{a} \hat{a}^2 \rangle \\
		\langle \hat{a}^\dagger \hat{a} \rangle & \langle    \hat{a}^\dagger \hat{a}^2 \rangle & \langle    \hat{a}^{\dagger 2} \hat{a} \rangle & \langle    \hat{a}^{\dagger 2} \hat{a}^2 \rangle \\
	\end{vmatrix}.
\end{equation}
As before, its associated four-copy observable $\hat{B}_{1235}$ can be obtained by assigning a different mode to each row and averaging over all $|S_4|=4!$ permutations, namely 
\begin{widetext}

\begin{equation}\label{def:d1235matrixBrun}
\hat{B}_{1235} = \frac{1}{|S_4|} \sum_{\sigma \in S_4}
\begin{vmatrix}
1 &     \hat{a}_{\sigma(1)} &    \hatd{a}_{\sigma(1)} &      \hat{a}^\dagger_{\sigma(1)} \hat{a}_{\sigma(1)} \\
    \hatd{a}_{\sigma(2)} &     \hatd{a}_{\sigma(2)} \hat{a}_{\sigma(2)} &  \hatdn{a}{2}_{\sigma(2)} & \hat{a}^{\dagger 2}_{\sigma(2)} \hat{a}_{\sigma(2)} \\
    \hat{a}_{\sigma(3)} &     \hat{a}^2_{\sigma(3)} &      \hatd{a}_{\sigma(3)} \hat{a}_{\sigma(3)} &      \hatd{a}_{\sigma(3)} \hat{a}^2_{\sigma(3)} \\
    \hat{a}^\dagger_{\sigma(4)} \hat{a}_{\sigma(4)} &     \hat{a}^\dagger_{\sigma(4)} \hat{a}^2_{\sigma(4)} &     \hat{a}^{\dagger 2}_{\sigma(4)} \hat{a}_{\sigma(4)} &     \hat{a}^{\dagger 2}_{\sigma(4)} \hat{a}^2_{\sigma(4)} \\
\end{vmatrix}.
\end{equation}

\end{widetext}
This expression of $\hat{B}_{1235}$ is lengthy but we know that calculating its mean value $\dmean{\hat{B}_{1235}}$ yields $d_{1235}$. 
Since it is an Hermitian operator, $\hat{B}_{1235}$ may be rewritten as $\hat{B}_{1235}=  ~ : \! \hatd{f}_{1235} \hat{f}_{1235} \! : $, where $\hat{f}_{1235}$ is Hermitian too and is defined as
\begin{widetext}
\begin{align}\label{eq:f1235operator}
\begin{split}
\hat{f}_{1235} = 
\frac{-i}{2 \sqrt{6}} &
(\hatd{a}_1 \hat{a}_1 \hatd{a}_2 \hat{a}_3 
- \hatd{a}_1 \hatd{a}_2 \hat{a}_2 \hat{a}_3
- \hatd{a}_1 \hat{a}_1 \hat{a}_2 \hatd{a}_3 
+ \hat{a}_1 \hatd{a}_2 \hat{a}_2 \hatd{a}_3
+ \hatd{a}_1 \hat{a}_2 \hatd{a}_3 \hat{a}_3 
- \hat{a}_1 \hatd{a}_2 \hatd{a}_3 \hat{a}_3 \\ &
- \hatd{a}_1 \hat{a}_1 \hatd{a}_2 \hat{a}_4 
+ \hatd{a}_1 \hatd{a}_2 \hat{a}_2 \hat{a}_4
+ \hatd{a}_1 \hat{a}_1 \hatd{a}_3 \hat{a}_4 
- \hatd{a}_2 \hat{a}_2 \hatd{a}_3 \hat{a}_4
- \hatd{a}_1 \hatd{a}_3 \hat{a}_3 \hat{a}_4
+ \hatd{a}_2 \hatd{a}_3 \hat{a}_3 \hat{a}_4 \\ &
+ \hatd{a}_1 \hat{a}_1 \hat{a}_2 \hatd{a}_4
- \hat{a}_1 \hatd{a}_2 \hat{a}_2 \hatd{a}_4
- \hatd{a}_1 \hat{a}_1 \hat{a}_3 \hatd{a}_4 
+ \hatd{a}_2 \hat{a}_2 \hat{a}_3 \hatd{a}_4
+ \hat{a}_1 \hatd{a}_3 \hat{a}_3 \hatd{a}_4 
- \hat{a}_2 \hatd{a}_3 \hat{a}_3 \hatd{a}_4 \\ &
- \hatd{a}_1 \hat{a}_2 \hatd{a}_4 \hat{a}_4
+ \hat{a}_1 \hatd{a}_2 \hatd{a}_4 \hat{a}_4
+ \hatd{a}_1 \hat{a}_3 \hatd{a}_4 \hat{a}_4 
- \hatd{a}_2 \hat{a}_3 \hatd{a}_4 \hat{a}_4
- \hat{a}_1 \hatd{a}_3 \hatd{a}_4 \hat{a}_4
+ \hat{a}_2 \hatd{a}_3 \hatd{a}_4 \hat{a}_4) ,\\
= \frac{2}{\sqrt{6}} & 
(\hat{L}_z^{12} \hat{L}_y^{34}  + \hat{L}_z^{13} \hat{L}_y^{42} + \hat{L}_z^{14} \hat{L}_y^{23} 
+ \hat{L}_y^{12} \hat{L}_z^{34}  + \hat{L}_y^{13} \hat{L}_z^{42}  + \hat{L}_y^{14} \hat{L}_z^{23}  ) ,\\ 
= \frac{1}{\sqrt{6}} & \sum_{\sigma \in P_4} \hat{L}_z^{\sigma (1) \sigma (2) } \hat{L}_y^{\sigma (3) \sigma (4) } ,
\end{split}
\end{align}
\end{widetext}
where $P_4$ is the group of even permutations. 
The expression of $\hat{f}_{1235}$ can be further simplified if we apply some linear optics transformation on the four modes. In analogy with the three-mode observable, we apply an orthogonal transformation on the mode operators that has the property of concentrating the coherent component of the four identical input states onto the first mode. By choosing the tensor product of two two-dimensional discrete Fourier transforms (realized with 50:50 beam splitters),
\begin{equation}
\begin{split}
\hat{a}'_1 &= \frac{1}{2}(\hat{a}_1+\hat{a}_2+\hat{a}_3+\hat{a}_4) , \\
\hat{a}'_2 &= \frac{1}{2}(\hat{a}_1-\hat{a}_2+\hat{a}_3-\hat{a}_4) , \\
\hat{a}'_3 &= \frac{1}{2}(\hat{a}_1+\hat{a}_2-\hat{a}_3-\hat{a}_4) ,\\
\hat{a}'_4 &= \frac{1}{2}(\hat{a}_1-\hat{a}_2-\hat{a}_3+\hat{a}_4) ,
\end{split}
\end{equation}
we obtain the simpler expression 
\begin{align}
\label{eq_final_expression_f_1235}
\begin{split}
\hat{f}_{1235} = -\sqrt{\frac{2}{3}}( \hat{L}_x^{23'} \hat{L}_y^{23'} + \hat{L}_x^{34'} \hat{L}_y^{34'} + \hat{L}_x^{42'} \hat{L}_y^{42'} ),
\end{split}
\end{align}
which only depends on modes 2, 3, and 4 as expected since $\hat{B}_{1235}$ is invariant under displacements. Thus,  $\hat{f}_{1235}$ is simply proportional to the scalar product between vectors  $(\hat{L}_x^{23'}, \hat{L}_x^{34'}, \hat{L}_x^{42'})^T$ and $(\hat{L}_y^{23'}, \hat{L}_y^{34'}, \hat{L}_y^{42'})^T$. However, we have not found a way to further simplify this expression and bring it to an experimental scheme. The problem is that $\hat{f}_{1235}$ is made of products of non-commuting operators $\hat{L}^{kl}_x$ and $\hat{L}^{kl}_y$, which, in addition, do not transform similarly when the mode operators undergo an orthogonal transformation. Unfortunately, applying local phase shifts and beam splitters does not help in reducing to an expression involving $\hat{L}_z^{kl}$ and $\hat{L}_0^{kl}$ only, as we were able to do for all previous multicopy observables.

Note that the issue does not seem to be related to the fact that we consider a four-copy observable. Indeed, while it is useless for nonclassicality detection, the principal minor $d_{25}$ can be expressed as the expectation value of the two-copy observable
\begin{equation}
\hat{B}_{25} = ~ : \! (\hat{L}_0 - \hat{L}_x)(\hat{L}_0^2-\hat{L}_z^2) \! : ~,
\end{equation}
which also cannot be accessed using only linear optics and photon number measurements.

Let us stress that the criterion based on $d_{1235}$ is in general stronger than those based on $d_{23}$ and $d_{15}$. In the special case of centered states, that is $\langle \hat{a} \rangle = \langle \hat{a}^\dagger \rangle = 0$, we show in Appendix \ref{sect_analysis_d_1235} that $d_{1235}$ can be negative even if $d_{15}$ and $d_{23}$ are both positive. However, if we further assume that $ \langle\hat{a}^{\dagger 2} \hat{a}  \rangle = \langle \hat{a}^{\dagger} \hat{a}^2\rangle = 0$, then we simply have
\begin{equation}
d_{1235} = d_{15} d_{23},
\end{equation}
in which case a separate implementation of the $d_{15}$ and $d_{23}$ circuits becomes sufficient to obtain the value of $d_{1235}$ and there is no need for a four-copy observable. This is the case for Fock, squeezed, and cat states.

\section{Conclusion and Perspectives}\label{sec:Conclusions}
\vspace{-0.2cm}

In summary, we have analyzed a family of nonclassicality criteria based on the matrix of moments of the optical field and (restricting ourselves to dimension $N=5$) have benchmarked their ability to detect nonclassical states, such as Fock states, squeezed states, and cat states. We have then developed a multicopy technique that allowed us to access these criteria by considering several identical replicas of the state and measuring the expectation value of some appropriate observables. For two- and three-copy nonclassicality observables, we have found a physical implementation that relies on linear optics and photodetectors with single-photon resolution. The main advantage of this multicopy technique is that it overcomes the need for full state tomography in order to detect nonclassicality. Furthermore, higher-order moments of the considered state can be accessed by measuring only lower-order moments such as the photon number on several replicas. The price to pay is of course the need to ensure interferometric stability on the replicas over which a joint observable is measured.

Specifically, we have found that the criteria based on  $d_{23}$ and $d_{15}$ are detecting well-known nonclassical features such as squeezing (for $d_{23}$) and sub-Poissonian photon number statistics (for $d_{15}$). These criteria can be accessed with a simple linear optics circuit applied on two replicas of the state, followed by photon number measurement. Then, we have found a stronger criterion based on   $d_{123}$, which is invariant under displacements but keeps the same detection performance as $d_{23}$. This criterion can be associated with a three-copy nonclassicality observable $\hat{B}_{123}$, which can again be accessed with interferometry and photon number measurements. Further, it also leads to a necessary and sufficient condition for the detection of nonclassicality in the entire set of Gaussian (pure and mixed) states.

Finally, we have identified the criterion based on $d_{1235}$ as the most remarkable of all criteria built from the minors of the matrix of moments of dimension $N=5$. It detects all states that are detected by $d_{15}$ and $d_{123}$ but also many more states that cannot be detected by neither of them. He have found a simple expression for its associated four-copy nonclassicality observable $\hat{B}_{1235}$, but have unfortunately not been able to find a corresponding physical implementation in terms of a linear interferometer and photon number detectors. Finding an implementation (involving probably a more complex circuit or higher-order measurements) is left as a challenging open problem. The difficulty arises because of the much higher order in mode operators of the observable $\hat{B}_{1235}$. Since we deal with three modes in Eq.~(\ref{eq_final_expression_f_1235}) rather than two modes as in Eq.~(\ref{eq:B123operator_1st_expression}), a possible approach could be to exploit the adjoint representation formed by $3\times 3$ matrices instead of the fundamental representation of SU(2)  (i.e., the set of $2\times 2$ Pauli matrices), written in terms of mode operators. Even though $d_{1235}$ only gives a sufficient (not necessary) condition for nonclassicality, it is expected to detect a wide range of nonclassical states, so that finding a feasible optical implementation of $d_{1235}$ would be highly valuable. Of course, in order to avoid false positives, one would have to assess the robustness of this nonclassicality condition against slightly different replicas, which will necessarily be the case in practice.

Another promising direction that is left for further work would be to extend our technique for detecting the nonclassicality of multimode optical states. We should be able to develop multicopy multimode observables, where each replica is made of several modes. Overall, this could result in an experimentally-friendly procedure to certify optical nonclassicality.


\begin{acknowledgements}
The authors warmly thank S. De Bi\`evre, F.~Grosshans, and T. Haas for useful discussions. C.G. is Research Fellow of the Fonds de la Recherche Scientifique – FNRS. N.J.C. acknowledges  support  by  the  Fonds de la Recherche Scientifique – FNRS  under  Grant No T.0224.18  and  by  the  European Union  under project ShoQC within ERA-NET Cofund in Quantum Technologies (QuantERA) program.
\end{acknowledgements}

\appendix

\section{Matrix of moments $D_N$}
\label{app:matrixofmomentsproperties}

\subsection{Block structure of $D_N$.}
\begin{figure}[b!]
	\begin{center}
		\includegraphics[width= 0.5\linewidth]{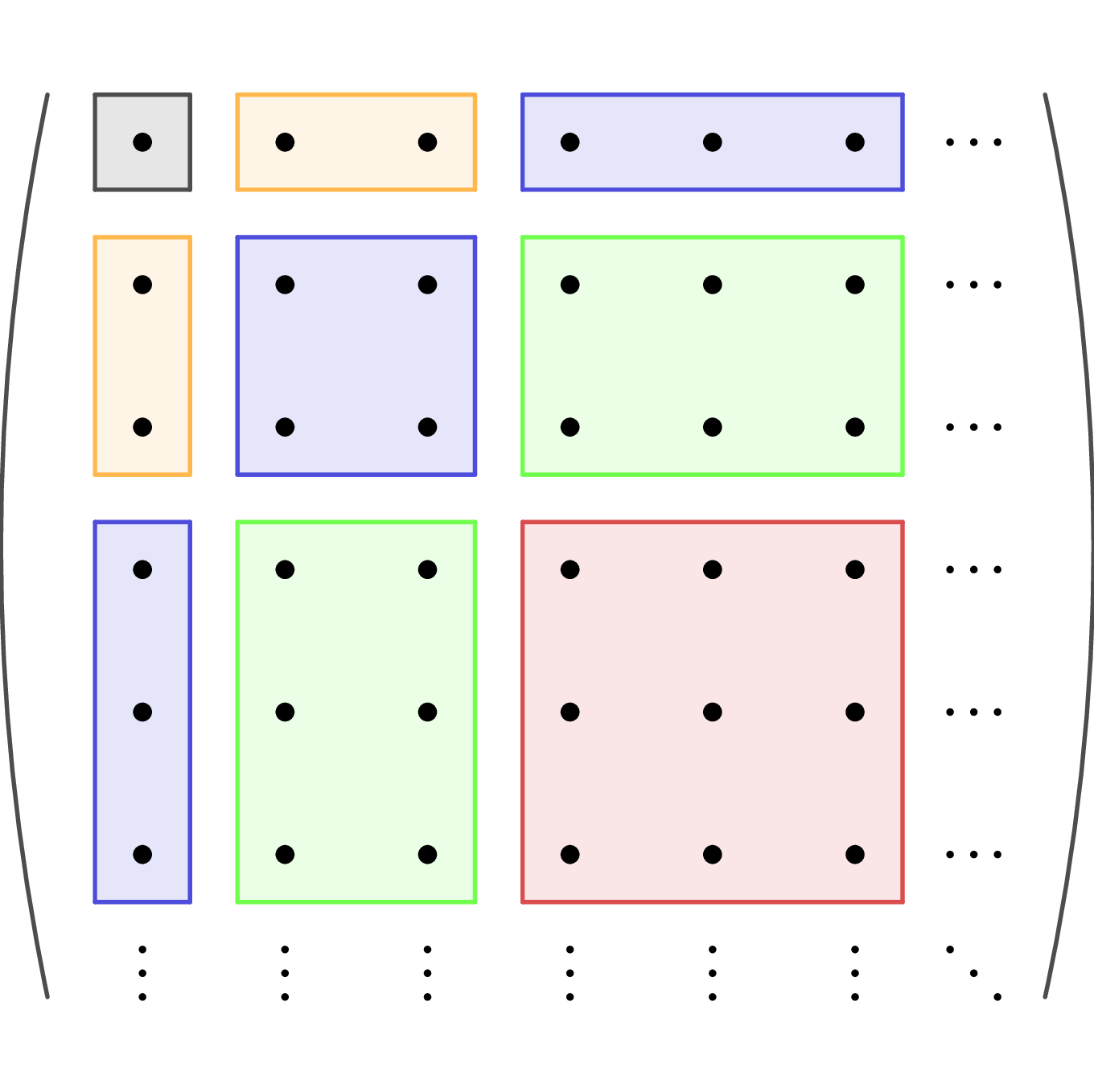}
		\caption{Block structure of the matrix of moments. The blocks of the same color are composed of entries having the same total power in $\hat{a}$ and $\hatd{a}$.}
		\label{fig:BlockStructure}
	\end{center}
\end{figure}

The matrix of moments $D_N$ is defined in Eq. (\ref{def:dNmatrix}).
By inspection, we see that the entry $m_{i,j}$ of row $i$ and column $j$ results from the (expectation value of the) product of the first entries in the same row and same column taken in normal order, namely  
\begin{equation}\label{eq:dnmatrixelements}
m_{i,j} = \mean{\hat{m}_{i,j}} = \mean{:\hat{m}_{i,1} \, \hat{m}_{1,j}:} =  \mean{:\hatd{m}_{1,i} \, \hat{m}_{1,j}:} ,
\end{equation}
where the entries of the first row are defined as
\begin{equation}\label{eq:dn1stlineoperators}
\hat{m}_{1,j} = \hatdn{a}{l} \hatn{a}{n-l}.
\end{equation}
In this expression, the index $j$ is decomposed into two indices $n$ and $l$ via 
\begin{equation}
j = \frac{n(n+1)}{2} + l + 1  ,
\end{equation}
which can be understood from the block structure of the matrix $D_N$ as illustrated in Fig. \ref{fig:BlockStructure}. The index $n$ stands for the block index and goes from 0 to infinity (when $N\to \infty$). It corresponds to the total order in $\hat{a}$ and $\hatd{a}$ of the block, while $l$ goes from 0 to $n$ corresponds to the order in $\hatd{a}$ (the order in $\hat{a}$ being of course $n-l$) within the $n$th block.

\subsection{Basic properties of $D_N$}

\paragraph{Hermiticity.}
This is easily seen from Eq. \eqref{eq:dnmatrixelements} since
$m_{i,j} = \mean{:\hatd{m}_{1,i} \, \hat{m}_{1,j}:} =  \mean{:(\hatd{m}_{1,j} \, \hat{m}_{1,i})^{\dagger}:} = m_{j,i}^*$.

\paragraph{Invariance under rotations.}
The invariance of $d_{1\cdots N} = \det(D_N)$ under rotations is easy to prove. Indeed, applying a rotation transforms the operator $\hat{a}$ into $e^{i \theta} \hat{a}$ and $\hatd{a}$ into $e^{-i \theta} \hatd{a}$. Since each term in the development of the determinant of $D_N$ always involves the same power in $\hat{a}$ and $\hat{a}^\dagger$, the factors $e^{i \theta}$ and $e^{-i \theta}$ cancel each other out. Hence, the observable is not affected by phase shifts and the criterion is invariant under rotations. For example, we have
\begin{align}
\begin{split}
d_{123}^{\theta} & = \begin{vmatrix}
	1 & \langle \hat{a} e^{-i \theta} \rangle & \langle \hatd{a} e^{i \theta}\rangle\\
	\langle \hatd{a} e^{i \theta}\rangle & \langle \hatd{a} e^{i \theta} \hat{a} e^{-i \theta} \rangle & \langle \hat{a}^{\dagger 2} e^{2i \theta}\rangle \\
	\langle \hat{a} e^{-i \theta}\rangle & \langle \hat{a}^2 e^{-2i \theta}\rangle & \langle \hatd{a} e^{i \theta} \hat{a} e^{-i \theta} \rangle  \\
\end{vmatrix}, \\
& = \mean{\hatd{a} \hat{a}}^2 - \mean{\hat{a}^{\dagger 2}} \mean{\hat{a}^2} 
- 2 \mean{\hatd{a}} \mean{\hat{a}} \mean{\hatd{a} \hat{a}} \\
& ~~~~~~ + \mean{\hat{a}^{\dagger 2}} \mean{\hat{a}}^2 
+ \mean{\hatd{a}}^2 \mean{\hat{a}^2},\\
& = d_{123},
\end{split}
\end{align}
where the superscript $\theta$ means that the state has been rotated by an angle $\theta$. In view of the form of $D_N$, it is clear that this rotation invariance also holds for all principal minors (and not just dominant principal minors). 

\paragraph{Invariance under displacements.}
The invariance under displacements of the dominant principal minors $d_{1\cdots N} = \det(D_N)$  of the matrix of moments can be understood from the simple example of $d_{123}$. It exploits a property of the determinant, namely that adding to a column (or row) a linear combination of any other columns (or rows) does not change the value of the determinant. By using this property recursively to $d_{123}$, we show that it is equal to $d^{\alpha}_{123}$ where the superscript $\alpha$ means that the state has been transformed by the displacement operator $\hat{D}(\alpha) = \exp{ \left( \alpha \hat{a}^\dagger - \alpha^* \hat{a} \right)}$, which transforms the operator $\hat{a}$ into $\hat{a}+\alpha$ and $\hatd{a}$ into $\hatd{a}+\alpha^*$.
\begin{widetext}
\begin{align}
\begin{split}
	d_{123} & = \begin{vmatrix}
		1 & \langle \hat{a} \rangle & \langle \hatd{a} \rangle\\
		\langle \hatd{a} \rangle & \langle \hatd{a} \hat{a} \rangle & \langle \hat{a}^{\dagger 2} \rangle \\
		\langle \hat{a} \rangle & \langle \hat{a}^2 \rangle & \langle \hatd{a} \hat{a} \rangle  \\
	\end{vmatrix}, \\
	& = \begin{vmatrix}
		1 & \langle \hat{a} \rangle +\alpha & \langle \hatd{a} \rangle + \alpha^\ast\\
		\langle \hatd{a} \rangle & \langle \hatd{a} \hat{a} \rangle + \alpha \langle \hatd{a} \rangle & \langle \hat{a}^{\dagger 2} \rangle + \alpha^\ast \langle \hatd{a} \rangle\\
		\langle \hat{a} \rangle & \langle \hat{a}^2 \rangle + \alpha \langle \hat{a} \rangle & \langle \hatd{a} \hat{a} \rangle  +\alpha^\ast \langle \hat{a} \rangle\\
	\end{vmatrix}, \\
	& = \begin{vmatrix}
		1 & \langle \hat{a} \rangle +\alpha & \langle \hatd{a} \rangle + \alpha^\ast\\
		\langle \hatd{a} \rangle + \alpha^\ast & \langle \hatd{a} \hat{a} \rangle + \alpha \langle \hatd{a} \rangle + \alpha^\ast (\langle \hat{a} \rangle +\alpha) & \langle \hat{a}^{\dagger 2} \rangle + \alpha^\ast \langle \hatd{a} \rangle + \alpha^\ast ( \langle \hatd{a} \rangle + \alpha^\ast)\\
		\langle \hat{a} \rangle + \alpha & \langle \hat{a}^2 \rangle  + \alpha \langle \hat{a} \rangle + \alpha(\langle \hat{a} \rangle +\alpha) & \langle \hatd{a} \hat{a} \rangle  +\alpha^\ast \langle \hat{a} \rangle + \alpha ( \langle \hatd{a} \rangle + \alpha^\ast)\\
	\end{vmatrix}, \\
	& = \begin{vmatrix}
		1 & \langle \hat{a}  +\alpha\rangle & \langle \hatd{a} + \alpha^\ast\rangle \\
		\langle \hatd{a}  + \alpha^\ast\rangle & \langle (\hatd{a}+\alpha^\ast) (\hat{a}+\alpha) \rangle & \langle (\hat{a}^{\dagger }+\alpha^\ast)^2 \rangle \\
		\langle \hat{a}  + \alpha \rangle & \langle (\hat{a}+\alpha)^2 \rangle  & \langle (\hatd{a}+\alpha^\ast) (\hat{a}+\alpha) \rangle \\
	\end{vmatrix},\\
	& = d_{123}^{\alpha}.
	\end{split}
\end{align}
\end{widetext}
This proof holds for any dominant principal minors $d_N$ but not for the other principal minors of the matrix of moments $D_N$. Remember that we adopt a slightly relaxed definition of dominant principal minors, which means that we must exhaust all rows and columns of a block before moving to the next block but the order of rows and columns is irrelevant within a given block.

\paragraph{Zero determinant for coherent states.}
Coherent states $| \alpha \rangle$ are extremal classical states. Indeed, any classical state can be written as a classical mixture of coherent states. Coherent states are of course not detected by all nonclassicality criteria since they are classical, but they come with a vanishing value for all principal minors (dominant or not) of the matrix of moment, which is consistent with extremality. Indeed, using $\hat{a} | \alpha \rangle = \alpha | \alpha \rangle$ and
		$\langle \alpha |\hat{a}^\dagger  =  \langle \alpha | \alpha^\ast$,
the matrix of moments can be written as
\begin{equation}
D^{\vert \alpha \rangle}_{123}  =  \begin{pmatrix}
	1 & \alpha &  \alpha^\ast \\
	\alpha^\ast & \alpha^\ast \alpha & \alpha^{\ast 2} \\
	\alpha & \alpha^2 & \alpha^\ast\alpha \\
\end{pmatrix}.
\end{equation}
One may see that $D^{\vert \alpha \rangle}_{123}$ is a rank-one matrix since it can be written as the outer product $\mathbf{\bar{\alpha}} \otimes \mathbf{\bar{\alpha}}^{\dagger}$, where $\mathbf{\bar{\alpha}} = (1, \alpha,  \alpha^\ast)^{\dagger}$. Hence, its determinant $d^{\vert \alpha \rangle}_{123}$ is equal to 0. In view of the form of the matrix of moments, this argument holds for all principal minors (of order strictly greater than one) which are therefore equal to 0 for coherent states. This shows the special role played by coherent states in the sense that they saturate all inequalities in Eq. (\ref{eq:CNSclassicality}).

\section{Jordan-Schwinger representation and\\ linear optics} \label{app:schwingerrepresentation}

The Jordan-Schwinger representation of the SU(2) algebra connects the angular momentum operators with the bosonic mode operators of two modes $\hat{a}_1$ and $\hat{a}_2$ as follows  
\begin{equation}
\hat{L}_j = \frac{1}{2} \hatd{A} \sigma_j \hat{A},
\end{equation}
where $ \hat{A} = (\hat{a}_1 \,\hat{a}_2)^T$    and $\sigma_j$ are the three Pauli matrices with $j = x, y, z$. The fourth associated operator is denoted as 
\begin{equation}
\hat{L}_0 = \frac{1}{2} \hatd{A} \sigma_0 \hat{A},
\end{equation}
where $\sigma_0 = \mathbf{1}$ is the $2 \times 2$ identity matrix, and it is linked to the Casimir operator $\hat{L}_x^2+\hat{L}_y^2+\hat{L}_z^2=\hat{L}_0(\hat{L}_0+1)$. We get a representation of angular momenta as we recover the usual commutation relations $[\hat{L}_j,\hat{L}_k]= i \,  \epsilon_{j,k,l} \, \hat{L}_l$ with $\epsilon_{j,k,l}$ being the Levi-Civita symbol, as well as 
$[\hat{L}_j,\hat{L}_0]= 0$, $\forall j$.

Let us derive the effect of linear optics transformations on the angular momentum operators. When applying a beam splitter of transmittance $\tau$, the angular momentum operators transform as
\begin{align}
    \begin{split}
       \hat{L}_0  & \rightarrow \hat{L}_0, \\
        \hat{L}_x  & \rightarrow  (1 - 2 \tau)~ \hat{L}_x + 2 \sqrt{\tau (1 - \tau)}~ \hat{L}_z, \\
         \hat{L}_y  & \rightarrow - \hat{L}_y, \\
          \hat{L}_z  & \rightarrow (2 \tau - 1)~ \hat{L}_z + 2 \sqrt{\tau (1 - \tau)}~ \hat{L}_x. \\
    \end{split}
\end{align}
When applying a phase shifter of phase $\phi$ on the second mode, the angular momentum operators transforms as
\begin{align}
    \begin{split}
       \hat{L}_0  & \rightarrow \hat{L}_0, \\
        \hat{L}_x  & \rightarrow  \cos(\phi)~ \hat{L}_x - \sin(\phi)~ \hat{L}_y, \\
         \hat{L}_y  & \rightarrow \cos(\phi)~ \hat{L}_y + \sin(\phi)~ \hat{L}_x, \\
          \hat{L}_z  & \rightarrow \hat{L}_z. \\
    \end{split}
\end{align}
These expressions are useful in order to find optical schemes for measuring the multicopy observables of interest. For example, we see that $\hat{L}_y$ transforms into $\hat{L}_x$ under a phase shift of $\pi/2$, while $\hat{L}_x$ transforms into $\hat{L}_z$ and vice versa under a 50:50 beam splitter transformation. This allows us to reexpress the multicopy observables of interest in such a way that they only depend on $\hat{L}_0$ and $\hat{L}_z$ operators (hence they are accessible via photon number measurement).

\section{Complementarity of $d_{15}$ and $d_{23}$}\label{app:complementarityexample}

As observed in Table \ref{table:determinantdeterminants} and Fig. \ref{fig:d1523detection}, it appears that $d_{15}$ and $d_{23}$ play complementary roles in the detection of nonclassical states. Indeed, the nonclassical pure states that we have studied are never detected simultaneously by $d_{15}$ and $d_{23}$, which is a consequence of Eq. (\ref{eq:complementarityd15d23}).
In order to illustrate this fact, we study an arbitrary superposition of Fock states $\vert 0 \rangle$, $\vert 1 \rangle$, and $\vert 2 \rangle$ with real amplitudes $a$, $b$, and $c$, namely,
\begin{equation}\label{def:012superposition}
\vert \psi_{012} \rangle =  a \vert 0 \rangle + b \vert 1 \rangle  +  c \vert 2 \rangle,
\end{equation}
where $a^2+b^2+c^2=1$.

\begin{figure}[t]
	\begin{center}
		\includegraphics[width=1\linewidth]{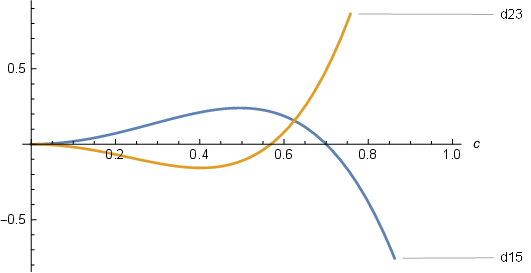}
		\caption{Comparison of the values of $d_{15}$ and $d_{23}$ for a ternary superposition of $\vert 0 \rangle$, $\vert 1 \rangle$, and $\vert 2 \rangle$ Fock states (with $b=0.1$). The criteria $d_{15}$ and $d_{23}$ are complementary in the sense that they do not simultaneously detect nonclassicality.}
		\label{fig:d15vsd23detection}
	\end{center}
\end{figure}

First, by setting $c=0$ in Eq. (\ref{def:012superposition}), i.e., for a superposition of $\vert 0 \rangle$ and $\vert 1 \rangle$ Fock states, we see that the determinant $d_{23}$ is always positive since $\mean{\hatdn{a}{2}} = \mean{\hatn{a}{2}} =0$, so the nonclassicality is not detected. In contrast, $d_{15}$ is always negative for all superpositions of $\vert 0 \rangle$  and $\vert 1 \rangle$ (this is expected since $d_{15}$ detects the nonclassicality of $\vert 1 \rangle$). The values of $d_{23}$ and $d_{15}$ are plotted in Fig. \ref{fig:d123VSd23}. Second, by setting $b=0.1$ in Eq. (\ref{def:012superposition}), the situation is a bit more complicated but it confirms the complementarity of $d_{23}$ and $d_{15}$, as shown in Fig. \ref{fig:d15vsd23detection}.

\section{Analysis of $d_{1235}$}
\label{sect_analysis_d_1235}

The criterion based on $d_{1235}$ is the strongest that we have found when considering the matrix of moments $D_5$. Although we have not found a multicopy optical scheme for accessing it, it is possible to simplify the problem by restricting to certain class of states. 
We consider here the case of centered states, which imposes some restrictions on the matrix of moment, namely that $\langle \hat{a} \rangle = \langle \hat{a}^\dagger \rangle = 0$.
In this case, the determinant $d_{1235}$ can be written as
\begin{equation}
	d_{1235} = \begin{vmatrix}
		1 &     0 &    0 &     \langle \hat{a}^\dagger \hat{a} \rangle \\
		0 &    \langle \hatd{a} \hat{a} \rangle&  \langle   \hatdn{a}{2} \rangle & \langle \hat{a}^{\dagger 2} \hat{a} \rangle\\
		0 &     \langle \hat{a}^2 \rangle & \langle     \hatd{a} \hat{a} \rangle & \langle     \hatd{a} \hat{a}^2 \rangle \\
		\langle \hat{a}^\dagger \hat{a} \rangle & \langle    \hat{a}^\dagger \hat{a}^2 \rangle & \langle    \hat{a}^{\dagger 2} \hat{a} \rangle & \langle    \hat{a}^{\dagger 2} \hat{a}^2 \rangle \\
	\end{vmatrix},
\end{equation}
which, by using the method of cofactors, simplifies to
\begin{equation}\label{eq:d1235factorized}
d_{1235} = d_{235} - \langle \hat{a}^\dagger \hat{a} \rangle^2 d_{23}.
\end{equation}
We may factorize the determinant $d_{235}$ by using the following property of determinants of block matrices.
If A and D are square matrices and if $A^{-1}$ exists, then
\begin{equation}\label{eq:detblockmatrix}
\det \begin{pmatrix}
A & B \\
C & D \\
\end{pmatrix}
= \det (A) \, \det (D - C A^{-1} B).
\end{equation}
Hence, assuming the extra constraint that $D_{23}$ is invertible, we get  
\begin{equation}\label{eq:d235factorized}
d_{235} = d_{23} \left(\langle \hat{a}^{\dagger 2} \hat{a}^2 \rangle - \begin{pmatrix}
	\langle \hat{a}^{\dagger} \hat{a}^2 \rangle & \langle \hat{a}^{\dagger 2} \hat{a} \rangle 
\end{pmatrix} D_{23}^{-1} \begin{pmatrix}
 \langle \hat{a}^{\dagger 2} \hat{a} \rangle \\
 \langle \hat{a}^{\dagger} \hat{a}^2 \rangle \\
\end{pmatrix} \right) .
\end{equation}
By plugging Eq. (\ref{eq:d235factorized}) into Eq. (\ref{eq:d1235factorized}), we rewrite $d_{1235}$ in a factorized form as
\begin{equation}\label{eq:d1235factorized1}
d_{1235} = d_{23} \left( d_{15} - \begin{pmatrix}
\langle \hat{a}^{\dagger} \hat{a}^2 \rangle & \langle \hat{a}^{\dagger 2} \hat{a} \rangle 
\end{pmatrix} D_{23}^{-1} \begin{pmatrix}
 \langle \hat{a}^{\dagger 2} \hat{a} \rangle \\
\langle \hat{a}^{\dagger} \hat{a}^2 \rangle \\
\end{pmatrix} \right).
\end{equation}
In particular, under the assumption that $D_{23}$ is positive definite (and hence $d_{23} >0$) and that $d_{15}<0$, then $d_{1235} <0$ and nonclassicality can be detected for sure. Moreover, we see that $d_{1235}$ is stronger than $d_{23}$ and $d_{15}$ since even if $d_{15}$ is positive but smaller than the second term in the right-hand side of Eq. (\ref{eq:d1235factorized1}), then $d_{1235}$ will detect the state as nonclassical.

An alternate decomposition of $d_{1235}$ can be obtained by exchanging the role of $d_{15}$ and $d_{23}$, which results in
\begin{equation}
\begin{split}
	d_{1235} &= d_{23} \, \det \left( D_{15} - \frac{1}{d_{23}} 
	\begin{pmatrix}
		0 & 0 \\
		0 & x
	\end{pmatrix} \right), \\
	&=  d_{23} \, d_{15}  - x.
\end{split}
\end{equation}
with $x=2 \langle \hat{a}^\dagger \hat{a} \rangle \langle\hat{a}^{\dagger 2} \hat{a}  \rangle \langle \hat{a}^{\dagger} \hat{a}^2\rangle - \langle \hat{a}^{\dagger 2} \rangle  \langle \hat{a}^{\dagger} \hat{a}^2\rangle^2 - \langle \hat{a}^{ 2} \rangle  \langle \hat{a}^{\dagger 2} \hat{a}\rangle^2$.
In this case again, we see that $d_{1235}$ is stronger than $d_{15}$ and $d_{23}$ as it can be negative even if $d_{15}$ and $d_{23}$ are both positive.

\bibliography{NonclassicalityMulticopyMeasurement_Article.bib}

\end{document}